\documentstyle[epsfig]{elsart}
\input astrobib.sty
\let\ov=\over
\let\l=\left
\let\r=\right

\def\be{\begin{equation}}
\def\ee{\end{equation}}

\def\spose#1{\hbox to 0pt{#1\hss}}
\def\lta{\mathrel{\spose{\lower 3pt\hbox{$\mathchar"218$}}
     \raise 2.0pt\hbox{$\mathchar"13C$}}}
\def\gta{\mathrel{\spose{\lower 3pt\hbox{$\mathchar"218$}}
     \raise 2.0pt\hbox{$\mathchar"13E$}}}

\begin{document}

\begin{frontmatter}

\title{Spectral methods in general relativistic astrophysics} 

\author{S.~Bonazzola, E.~Gourgoulhon \& J.-A.~Marck}

\address{D\'epartement d'Astrophysique Relativiste et de Cosmologie, \\
  (UPR 176 du C.N.R.S.), \\
	Observatoire de Paris, Section de Meudon, \\
	F-92195 Meudon Cedex, France \\
   {\em e-mail : 
	Silvano.Bonazzola,Eric.Gourgoulhon,Jean-Alain.Marck@obspm.fr} }

\begin{abstract}
We present spectral methods developed in our group to solve 
three-dimensional partial differential
equations. The emphasis is put on equations arising from astrophysical 
problems in the framework of general relativity.
\end{abstract}
\end{frontmatter}

\section{Introduction} \label{s:intro} 

Astrophysical problems involve many field of physics: hydrodynamics,
magneto hydrodynamics, chemistry, kinetical, nuclear and statistical
physics, gravitation... Physicists and astrophysicists have developed
well suited techniques to solve numerically mathematical problems in
each of these specific fields. Moreover they have to handle problems in
which different branches of physics interact and therefore
compatibility of different numerical techniques is required.

The following examples illustrate the above statement. Chaotic chemical
reactions can occur in the interstellar medium \cite{LebPF93}. Moreover
the interstellar medium is turbulent and mildly supersonic.  Therefore
the numerical modelizations requires numerical methods able to describe
chaos {\em and} compressible turbulence.  Astrophysicists involved in
numerical work are confronted with problems that require the integration
of systems of quasi-linear partial differential equations (PDEs) (i.e.
PDEs which are linear in the highest order derivatives). In general these
systems of equations contain all kind of PDEs {\em i.e.} hyperbolic,
parabolic and elliptic PDEs. The study of stellar oscillations is
another good example. The system of PDEs that must be integrated is the
Navier-Stokes equation for the fluid motion coupled to the continuity
equation for the mass conservation, to the scalar Poisson equation for
the gravitational field and to the heat equation (since the plasma of
the star has a finite heat conductivity).

PDEs systems for special relativistic astrophysics present the same
structure than the Newtonian ones but special relativistic hydrodynamic
equations are more complicated because of the presence of high Lorentz
factors which make life less easy. Relativistic jets in active galactic nuclei
are the realm of special relativistic [magneto-] hydrodynamics codes.
Things change drastically for compact astrophysical objects
such as neutron stars and black holes. For these objects 
the Newtonian theory of the gravitation is no longer appropriate and general
relativity must be used.

Differential geometry is the realm of GR. New geometrical concepts such as
topology, globality, isometry, Lie derivative, Killing vector,
curvature, etc... play an essential role. Of course all
these concepts are present in the framework of special relativity,  but
in a trivial and underlying way. Therefore, astrophysicists dealing
with special relativity, as a modern Monsieur Jourdain  who {\em ``fait
de la prose sans le savoir''} (he  speaks in prose without knowing
it), use these concepts without realizing it.

In general relativity, the gravitational potential is no longer scalar.
The metric tensor $g_{\alpha \beta}$, ($\alpha,\beta =
0,1,2,3$) replaces it, its 10 independent components playing the role
of gravitational potentials. The gravitational field equations 
(Einstein equations)
can be written in a beautiful very compact form:
\be \label{e:Einstein} 
	R_{\alpha \beta} - \frac{1}{2}Rg_{\alpha\beta}=
	{8\pi G \over c^4} \, T_{\alpha\beta} \ , 
\ee
where $R_{\alpha\beta}=R_{\beta \alpha}$  and $R$ are respectively the
Ricci tensor and the curvature scalar $R={R^{\alpha}}_{\alpha}$ 
associated with the metric $g_{\alpha\beta}$, and
where $T_{\alpha\beta}$ is the energy-momentum tensor. The equations of
motion can be written in a compact way too~:
$\nabla_{\alpha}{T^{\alpha}}_{\beta}=0$. If an electromagnetic field is
present, the Maxwell equations must be added to the above
equations~:  $\nabla_{\alpha}{F^{\alpha}}_{\beta} =J_{\beta}$ where
${F^{\alpha}}_{\beta}$ is the electromagnetic tensor and $J_{\beta}$ is
the quadri-electric current. The simplicity of these equations is only
apparent. In fact, when explicited, they contain a few
thousand of terms. Finding a solution to a complete 3-D physical
problem in the framework of general relativity is the {\em Holly Grail}
of Numerical Relativity.

Einstein equations form a system of $10$ second order
quasi-linear PDEs.  Moreover, the  $4$ identities (Bianchi
identities)
\be \label{e:Bianchi} 
	\nabla_{\alpha}({R^{\alpha}}_{\beta}
		- {1 \over 2} R {g^{\alpha}}_{\beta})
	\equiv 0
\ee 
introduce $4$ degrees of freedom for the solution. This is equivalent
to the freedom of gauge choice in electromagnetism. The
choice of the gauge in GR defines the character (i.e. ellipticity, parabolicity
or hyperbolicity) of the equations. 
Actually, the Einstein's equations are degenerated in the sense
that their character depends on the gauge
choice. For example, these equations are all
hyperbolic when the so-called {\em harmonic gauge} \cite{AbrY97} is chosen 
 and, on
the contrary, the system can be reduced to $5$ elliptic
equations coupled to $5$ hyperbolic equations if the so-called {\em
radiation gauge} \cite{SmaY78} is chosen.

The first problem which arises naturally when one wants to solve the
Einstein equation for a particular astrophysical problem is the
determination of the most appropriate gauge. There is no definitive
answer to this question and the debate is still open. One has to
consider essentially two criteria. The first one depends on the physics
of the problem: depending on the coordinate
system the space-time slicing induced by the gauge may or
may not be able to describe and cover the physical event that one wants
to study. The second one depends on the numerical technique employed:
depending on the gauge one may or may not be
able to find a numerical solution to the equations (e.g. ability
to solve coupled non-linear elliptic equations). It is tempting to
choose a gauge which seems more appropriate to the numerical technique
used. However, we don't think that this is a good approach to the
problem.

We hope to have given to the reader a vague idea of the kind of
problems met by astrophysicists and of the mathematical difficulties
that have to be overcome. We will show in this paper how spectral
methods may be used to successfully overcome these difficulties.
Spectral methods (SM) will be introduced in details in 
Sect.~\ref{s:ints}. For now, let us anticipate two of the most important
advantages of SM. The first one is the economy of the
number of degrees of freedom (NDF). For a given numerical accuracy,
less degrees of freedom are required than in the case of finite
difference techniques. More precisely, as a rule of thumb, we can say
that with SM, the number of grid points is reduced by a factor of
five per space dimension. This means that, in the case of 3-D problems,
the number of grid points will be decreased by a factor of
$5^{3}=125$.  Now, if one consider a dynamical problem where the
integration time step scales at least as $N$, where $N$ is the NDF,
this advantage becomes obvious. The second advantage consists
in the possibility of using a coordinate system well adapted to the
geometry of the problem and to handle exactly the pseudo-singularities
potentially present in the chosen coordinate system. For instance,
thanks to SM it has been possible to study 3-D
turbulent motion of an incompressible fluid with Reynolds number of
about $1000 $ \cite{VinM94} and to describe the disruption of a star in
the tidal field of a black hole in less than $3$ hours of C.P.U. time
on a workstation \cite{MarLB96} when the same problem, using a finite
difference method, has required $30$ hours of C.P.U. time on a Cray-2
\cite{KokNP93}. Moreover, thanks to the high precision achievable with
SM the stability and collapse of a N.S. close to its critical mass
was successfully studied in the GR frame \cite{Gou91}, \cite{GouHG95}.

Elliptic and parabolic PDEs are the realm of SM. Nevertheless good
results can be obtained for hyperbolic equations too if no
discontinuities appear in the solution \cite{BonM90}, \cite{MarLB96}.
When a discontinuity is present, some kind of natural
or artificial viscosity must be added in order to obtain a sufficiently smooth
solution. Interesting results were obtained with this technique in
describing a mildly supersonic turbulence occurring in star formation
\cite{BonHP87,PasP88,VasPP95,PasVP95}, or, by our group, in
modelling a 3-D stellar collapse \cite{MarB92}. 
Other numerical methods (especially those based on Riemann
solvers technique \cite{Mar97}) have been developed in order to handle shocks.
Successful results were obtained in modelizing relativistic jets (see
\cite{Iba98} and literature quoted there). In the case of
relativistic problems where systems of elliptical coupled to hyperbolic
PDEs must be solved, an interesting strategy consists in using SM
for the elliptical equations and Riemann solver technique for
the hyperbolic equations where discontinuities may occur. Interesting
results have already been obtained in 1-D modellisation of the collapse
of a core of a supernova in the framework of a tensor-scalar theory of
gravitation \cite{Nov98a}, \cite{Nov98b}. This strategy looks very
promising too for 3-D problems.

Our aim in the present paper is to explain how SM work and how they can
be implemented in the easiest way. The mathematical theory of SM will
be left to specialized literature (e.g. \cite{GotO77},
\cite{CanHQ88}). The reader does not need to know astrophysics but
almost all the examples presented as illustration are taken from the
modelization of astrophysical phenomena. Therefore some short
explanation of the phenomenon that has motivated the modelization will
be given. On the contrary some elementary notions of differential
geometry would be worth for understanding the mathematical problems.

The plan of this paper is as follows. We give first a short introduction to
spectral methods (Sect.~\ref{s:ints}) in which the essential notions are 
presented. Then we explicit the algebraic properties of representation of some
relevant differential operators (Sect.~\ref{s:algeb}). Basic methods
for solving PDEs are presented in 
Sect.~\ref{s:solution_PDE}, especially the time scheme and the treatment of
boundary conditions. This is illustrated by selected one-dimensional examples 
(Sect.~\ref{s:Exa}). Three-dimensional (3-D) 
spectral methods with spherical-like coordinates are introduced in 
Sect.~\ref{s:3d}. Scalar and vectorial equations, such as Poisson or
telegraph equations, are considered in Sects.~\ref{s:poisson-scal} and
\ref{s:veceq} respectively. After a brief discussion 
of the structure of Einstein equations,
the paper ends by the presentation of various astrophysical results 
(Sect.~\ref{s:astro}).


\section{Introduction to spectral methods}\label{s:ints}

\subsection{A simple example based on a Fourier expansion}\label{s:fexp}

In this section we present the main ideas underlying spectral
methods (SM).  We shall give only the basic notions, living the theory to
specialized and very good existing text books, among which we recommend the
monographies by Gottlieb \& Orzsag \cite{GotO77} and by Canuto, Hussaini
Quarteroni \& Zang \cite {CanHQ88}.

Let us consider the quasi-linear partial differential equation 
(viscous Burger equation):
\begin{equation}\label{e:bour}
{\partial u\over\partial t} = {\partial^2 u\over \partial x^2} 
	+ \lambda u {\partial u\over \partial x} \ ,
	\quad t \ge 0 \ , \quad x \in [0,1] \ ,
\end{equation}
where $u$ is a function of the two variables $t$ and $x$ and $\lambda$
is some parameter. 

Consider first the simple case $\lambda =0$ (linear heat equation). 
Let us assume that the
value of $u(t,x)$ is known for all $x$ at $t=0$ and that $u$ is a
periodic function. The basic idea underlying SM consists
to transform the PDE in a system of ordinary differential equations by
means of a expansion of the solution onto a series on a complete
basis. Since $u$ is assumed to be periodic, it is natural to use a Fourier
expansion:
\be\label{e:esfou}
	u(x,t) = \sum_{k=0}^{\infty} \left[
	  a_{k}(t) \cos\left( 2\pi k x \right)
		+ b_{k}(t) \sin\left( 2\pi k x \right)\right] \ .
\ee
Equation.~(\ref{e:bour}), with $\lambda=0$, can then be rewritten as
\be \label{e:boufu}
\frac{da_{k}}{dt}=-k^{2}a_{k}(t)\ , \quad \frac{db_{k}}{dt}=-k^{2}b_{k}(t) \ .
\ee
In this way, finding a solution of the PDE (\ref{e:bour}) turns out to
be equivalent to solve the infinite system of ordinary differential
equations (\ref{e:boufu}), where the initial values of the coefficients
$a_{k}$ and $b_{k}$ are given by the Fourier expansion of the function $u$
at the time $t=0$:
\be\label{e:coeff}
a_{k}(0)=\int_0^1 u(x,0)\cos(2\pi k x)\, dx,\qquad b_{k}(0)=\int_0^1 u(x,0)\sin
(2\pi k x)\, dx \ .
\ee
From the numerical point of view, the series (\ref{e:esfou}) has to be
truncated. Let be $N/2-1$ the highest term of the series ($N$ even).
The integrals (\ref{e:coeff}) must be evaluated numerically in the
most accurate way: if $u(x,0)$ does not contain spatial frequencies
higher than $N/2-1$, then its Fourier  coefficients $a_{k}$ and $b_{k}$
must be computed exactly within the roundoff errors. In other words we
require that the numerical integrals
\begin{eqnarray} 
& & \int_{0}^{1} \cos(2\pi kx)\cos(2\pi jx)dx \ ,
	\int_{0}^{1} \sin(2\pi kx)\sin(2\pi jx)dx \ , \nonumber \\
& & 	\int_{0}^{1} \sin(2\pi kx)\cos(2\pi jx)dx   \label{e:num}
\end{eqnarray}
are computed exactly. The minimum number of grid points at which the
functions $u(x,0)$ must be sampled is $N$. The sampling points that
fulfil the above requirements are called the {\em Gauss-Lobatto points} or
{\em collocation points}: $x_{j}=j/N$. These points are equally
spaced\footnote{The properties of the Gauss-Lobatto sampling for the
discrete Fourier decomposition can be recovered easily by bearing in
mind that $\sum_{n=0}^{N-1}q^{n}=(1-q^{N})/(1-q),\ \ |q| \leq 1$, from
which the following identity holds ($q=\exp (2\pi i(k-j)/N)$):
\begin{eqnarray}\label{e:ident}
   N\,\int_{0}^{1}\exp (2i\pi (k-j) x)dx & = &  
		\sum_{n=0}^{N-1}\exp(2i\pi (k-j) x_{n})  \nonumber \\
 & = & \sum_{n=0}^{N-1}\exp (2\pi i(k-j)n/N)=N\delta_{k\,j}
\end{eqnarray}
This means that the computation of the integrals (\ref{e:num}) by means of
the trapeze method is exact (within the
round-off errors).  In the theory of signal, the Gauss-Lobatto
points correspond to the Shanonn sampling, and $N/2-1$ is related to the
Shanonn frequency.}.

Note that there exists an isomorphism between the coefficients
$a_{k}(t),b_{k}(t)$ and the values $u(x_{j},t)$ of the function at the 
collocation points. For this
reason, $N$ is called the {\em number of degrees of freedom} (NDF),
denomination that
we prefer to the more usual {\em number of grid points}. Finally it is
worth to note that the uniform repartition of the Gauss-Lobatto points 
should be considered as an accident. If one changes the basis of
expansion, uniform spacing does not  a priori  correspond to a
Gauss-Lobatto sampling (see the next section). 

Consider now the non linear problem $\lambda \neq 0$. We can proceed in
two different ways to treat the non-linear term. The first one 
({\em spectral method} in the strict sense) 
consists in computing the first space derivative of $u$
in the Fourier space (or coefficient space) and to perform a
convolution with $u_{k}$. From a numerical point of view, these terms,
computed at the time $t^{j}$, act as a source and allow to compute the
coefficient $u_{k}$ at the time $t^{j+1}$ no matter the temporal scheme
used. Within spectral method, all the evolution of the equation is
computed in the Fourier space. Backing in the configuration space (at
arbitrary values of $x$) is performed only for practical convenience, for
example to visualize the solution at a given time. Consequently, the
configuration space plays a secondary role.

The second way to proceed {\em (pseudo-spectral method)} consists in computing
the derivative of $u$ in the Fourier space, to come back in the
configuration space by an inverse Fourier transform, to multiply
$\partial u/\partial x$ by $u$ in the configuration space and to come
back again in the Fourier space.  With pseudo-spectral method, the Fourier
and configuration spaces behave equally. Pseudo-spectral method is
generally used in spite of the apparent athletic work of dancing from
one space to the other. The reason is that performing a convolution
requires a number of operations proportional to $N^{2}$. On the
contrary, thanks to the fast\footnote{A 1-D numerical algorithm is
called {\em fast} if the number of arithmetic operations scales
with the number of the degrees of freedom $N$ no faster than $N\,\log
N$. Fast Fourier Transform (FFT) and Fast Chebyshev transform
algorithms (FCT) are {\em fast} algorithms.} Fourier algorithm, the
number of operations needed by pseudo-spectral method is proportional
to $N \log N$. Moreover the pure spectral method can not handle easily
non-linearities more general than the quadratic one present in
(\ref{e:bour}). In all this paper, we shall use (incorrectly) for
shortness the term of {\em spectral method} instead of the more
appropriate {\em pseudo-spectral method}.

The above example contains all the philosophy of SM.
A PDE, or more generally a system of PDEs, is transformed in an
equivalent system of ordinary differential equations. The set of
trigonometric functions used for the expansion, has been chosen because
it fulfils automatically the boundary conditions and because it exists
a fast transform algorithm.  The space derivatives are computed in the
coefficients space and the computation uses the value of the function
on {\it all} collocation points. Quadratic terms are computed by
coming back to the configuration space. 
The most important consequence is that
if the solution is $C^{\infty}$ then the ${\cal L}^{2}$ error between the
numerical solution and the analytical one scales as $\exp(-N)$ and as
$N^{-(p+1)}$ for a solution $C^{p}$\footnote{Following the terminology
of the functional analysis, the error ${\cal L}^{p}$ of some numerical solution
$u_{\rm num}(x,t)$ is defined by
\be
 \left( \int_{-1}^{1}\left [u_{\rm num}(x,t)-u_{\rm ex}(x,t)\right]^{p}dx
	\right)^{1/p}
\ee
where $u_{\rm ex}(x,t)$ is the exact solution.}.
An error that scales as $\exp(-N)$ is called {\em evanescent}. 

Finally the computation of the required quantities is performed in one
of the two spaces, namely the configuration space or its dual
(Fourier space), in order to save computation time and to obtain high
accuracy.

\begin{figure}
\centerline{\epsfig{figure=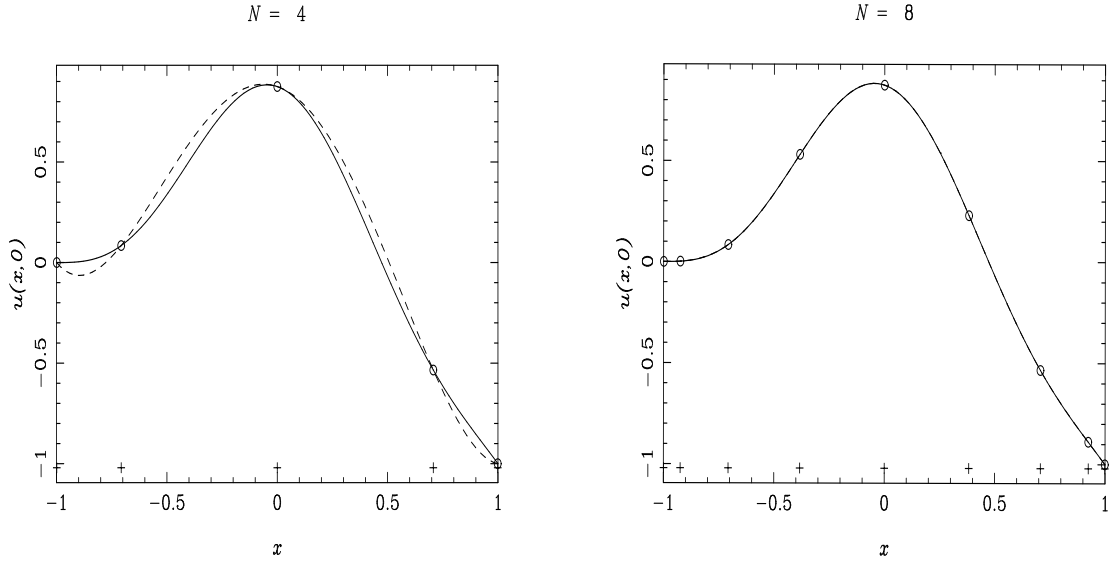,height=8cm}}
\caption[]{\label{f:u,I4,I8}
Graph of the function $u(x) = \cos^3 ( \pi\, x/2) - (x+1)^3 /8$ 
(solid line) and its Chebyshev interpolant (dashed line)
for $N=5$ (left) and $N=9$ (right). For $N=9$, the two curves cannot be
distinguished graphically. The circles denote the values at the
collocation points $x_n$. Note that spurious oscillations between the 
collocation points are not present. 
}    
\end{figure}

\begin{figure}
\centerline{\epsfig{figure=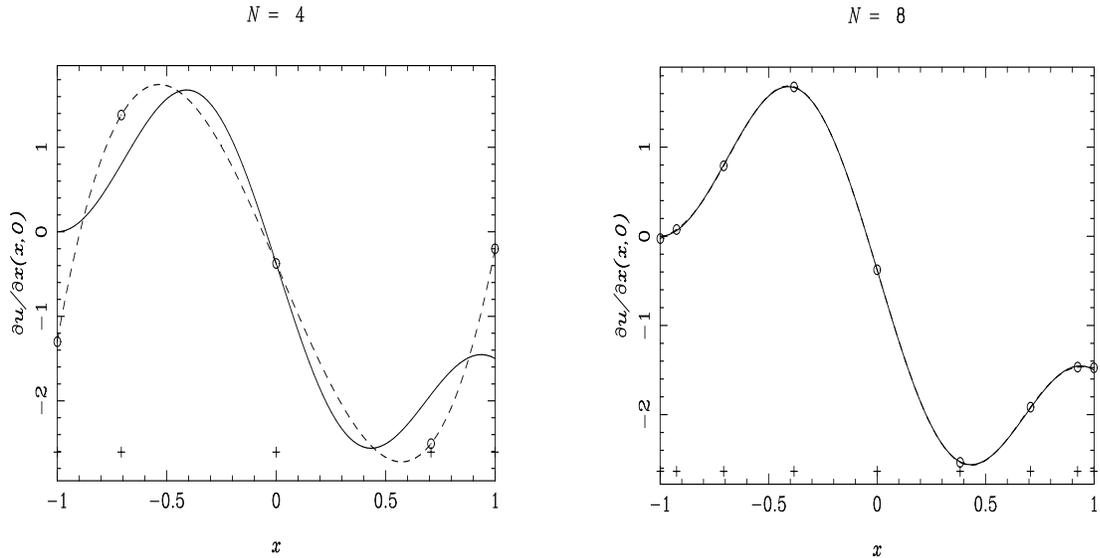,height=8cm}}
\caption[]{\label{f:dudx,I'4,I'8}
Graph of the derivative of the function represented in Fig.~\ref{f:u,I4,I8}
Eq.~(\ref{e:u=cos3}) (solid line) and its Chebyshev approximation (dashed line), 
for $N=5$ (left) and $N=9$ (right). The circles denote the values at the
collocation points $x_n$.}    
\end{figure}

\subsection{More general expansions}

If the initial conditions or the boundary conditions are not periodic,
Fourier expansion is no more adapted to treat the problem because of
the presence of a Gibbs phenomenon at the boundaries of the interval
\footnote{The Gibbs phenomenon occurs when an expansion, for example
the Fourier expansion, of a discontinuous function is
performed. In this case the ${\cal L}^{2}$ error scales with the
degrees of freedom $N$ as $1/N$. On the contrary the pointwise error
${\cal L}^{\infty}$ tends to a finite value when $N \to \infty$
(see e.g. ref.~\cite{GotO77})}. A new base of functions must be used. Expansions in
Legendre or Chebyshev polynomials are the most common ones. The reader
will find in \cite{GotO77} and \cite{CanHQ88} a discussion on the
advantage and drawbacks of the different expansions. Legendre
polynomials are often used for theoretical work \cite{MadOT93}. The
main advantage of Chebyshev polynomials is that there exists a fast
algorithm to perform Chebyshev expansion (its C.P.U. time is just a
little bit longer than the corresponding Fourier transform time). For
this reason, Chebyshev expansion is widely used in this paper. 

{\em Chebyshev polynomials} $T_{n}(x)$ are defined as the family of polynomials
which obeys to the the following orthogonality relations:
\be\label{e:Cheb} 
{ (1+\delta_{n\,0}) \over \pi}
\int_{-1}^{1} {T_{n}(x)T_{m}(x) \over \sqrt{1-x^{2}} } dx
	= \delta_{n\,m}  \ , 
\ee
where $\delta_{n\,m}$ is the Kronecker symbol.  By introducing the new
variable $\theta=\arccos(x)$ ($0\leq \theta \leq \pi$), the Chebyshev
polynomials read
\be\label{e:Chebt}
T_{n}(x(\theta ))=\cos(n\theta ) \ .
\ee
The associated Gauss-Lobatto points $x_{j}$ are uniformly spaced with
respect to the variable $\theta$: $x_{j}=\cos(\pi(j-1)/(N-1)),\ \ 0
\leq j \leq N-1$ where $N$ is the number of degrees of freedom. It easy
to see that the collocation points are more dense close to the
boundaries of the interval $[-1,+1]$. Thanks to the relation (\ref{e:Chebt}),
Chebyshev polynomials have close links to the trigonometric functions used
in the Fourier expansion.

The following example illustrates the accuracy of SM.
Let us consider the function
\be \label{e:u=cos3}
   u(x) = \cos^3 \l( {\pi\ov 2} x \r) - {1\ov 8} (x+1)^3 \qquad
	x \in [-1,1] \ .  
\ee
Note that $u(x)$ is not a polynomial, so that its expansion over the
Chebyshev polynomials a priori requires an infinite number of coefficients.
The term $(x+1)^3$ has been added to the cosine term in order not to have a 
periodic function.  
The function $u(x)$ is represented in 
Fig.~\ref{f:u,I4,I8}, as well as its Chebyshev interpolant 
$\sum_{n=0}^{N-1} \hat u_n \, T_n(x)$ for
$N=5$ and $N=9$. Note that for a number as small as $N=9$, the Chebyshev
expansion cannot
be graphically distinguished from $u$. The derivative of $u(x)$ is
numerically computed by means of the method described in Sect.~\ref{s:der}
and is compared with its analytical value in Fig.~\ref{f:dudx,I'4,I'8}. 

The comparison with a second-order finite difference method is performed in
Table~\ref{t:erreur} for the computation of the first and second derivatives
of $u$. Note
that for $N=33$, the SM reaches a relative accuracy of $1.5\times 10^{-13}$ and
$3\times 10^{-11}$ in the computation of $du/dx$ and $d^2u/dx^2$ respectively.

\begin{table}
\begin{center}
\begin{tabular}{rcccc}
\hline
NDF 	& \multicolumn{2}{c}{error on $\displaystyle {du\over dx}$} 
	& \multicolumn{2}{c}{error on $\displaystyle {d^2u\over dx^2}$} \\
	& SM & FDM & SM & FDM \\
\\ \hline
      	5 &  5.8E-01  & 8.2E-02   & 2.1E+00  & 3.7E+00 \\     
        7 &  1.2E-01  & 5.5E-01	  & 3.4E-01  & 2.2E+00 \\   
        9 &  1.2E-02  & 5.1E-01   & 5.8E-02  & 1.4E+00 \\    
       13 &  2.8E-05  & 2.6E-01	  & 3.6E-04  & 6.8E-01 \\   
       21 &  4.2E-12  & 1.0E-01   & 1.6E-10  & 2.6E-01 \\     
       33 &  1.5E-13  & 4.0E-02	  & 2.9E-11  & 1.0E-01 \\    
       65 &  3.2E-13  & 1.0E-02   & 5.6E-10  & 2.6E-02 \\
      129 &  2.5E-11  & 2.5E-03   & 1.4E-07  & 6.4E-03 \\
      513 &  1.4E-09  & 1.6E-04	  & 8.6E-05  & 4.0E-04 \\
     1025 &  1.6E-08  & 4.0E-05	  & 5.3E-03  & 1.0E-04 \\
\hline
\end{tabular}
\end{center}
\caption{\label{t:erreur}
Comparison between a Chebyshev spectral method (SM) and a
second-order finite difference method (FDM) in the computation of the 
first and second derivatives of the function 
$u(x) = \cos^3 \l( {\pi\ov 2} x \r) - (x+1)^3/8$. The error is defined as the highest discrepancy with
the exact value at the grid points. 16 digits are used in the
numerical computation. Note that to obtain an accuracy on $d^2u/dx^2$ of the
order $10^{-4}$, 13 points are sufficient for the SM, whereas the FDM requires
about 500. Note also that for  
${\rm NDF}\geq 129$, the round-off errors make the SM less efficient. }
\end{table}
\section{Algebraic properties and spectral representation of some elementary 
operators}\label{s:algeb}

We will discuss in this section some properties and implementation of
elementary linear differential operators when using Chebyshev polynomial
expansions. 

\subsection{First and second order derivatives.}\label{s:der}

The expansion coefficient  $b_i$ of $d f / d x$ are given by $b_i =
D_{ij} a_j$ where $a_i$ are the coefficients of $f$ and, 
by definition of $d/dx$,
\be
\label{eq:matd1a}
 	D_{ij}  =  (2-\delta_0^i) \int_{-1}^{1}T_{i}(x)
			{T'}_{j}(x)(1-x^{2})^{-1/2}dx \ .
\ee
The matrix $D_{ij}$ is computed from the relation
\be
   {T'}_{n+1}(x) = 2(n+1)\, T_n(x) + {n+1\ov n-1} {T'}_{n-1}(x) \ ,
\ee
which can be derived easily from Eq.~(\ref{e:Chebt}).  
For a NDF = 7, the result is:
\def\arraystretch{.9}
\be \label{e:mat(d/dx)}
\arraycolsep=3mm
   D_{ij} = \left[ \begin{array}{ccccccc}
    0 & 1 & 0 & 3 & 0 & 5 & 0  \\
    0 & 0 & 4 & 0 & 8 & 0 & 12 \\
    0 & 0 & 0 & 6 & 0 &10 & 0  \\
    0 & 0 & 0 & 0 & 8 & 0 & 12 \\
    0 & 0 & 0 & 0 & 0 &10 & 0  \\
    0 & 0 & 0 & 0 & 0 & 0 & 12 \\
    0 & 0 & 0 & 0 & 0 & 0 & 0 
   \end{array} \right] \ .
\ee 
The disposition of the non vanishing elements of $D_{ij}$ is due to
the property of the operator $d/dx$  which transforms the parity of the
functions at which is applied. The main diagonal and the last line of
$D_{ij}$ vanish because the operator $d/dx$ transforms a polynomial
of degree $n$ into a polynomial of degree $n-1$. These two important
properties will be used below.

The representation of the operator $d^2 / dx^2$ can be obtained
straightforwardly by taking the square of the matrix
(\ref{e:mat(d/dx)}). For a NDF $= 7$, it reads 
\def\arraystretch{.9}
\be \label{e:Mat2}
\arraycolsep=3mm
D^{2}_{ij}=\left[\begin{array}{rrrrrrrrr} 
      0&     0&     4&     0&    32&     0&    108\\          
      0&     0&     0&    24&     0&   120&     0\\
      0&     0&     0&     0&    48&     0&    192\\
      0&     0&     0&     0&     0&    80&     0\\
      0&     0&     0&     0&     0&     0&    120\\
      0&     0&     0&     0&     0&     0&     0\\
      0&     0&     0&     0&     0&     0&     0
\end{array}
\right]
\ee
The disposition of the zeros is different, because the operator
$d^{2}/dx^{2}$ preserves the parity. The last two lines of the matrix
vanish because this operator decreases the  degree of a polynomial by
two units. Other operators that preserve the number of coefficients are
\be\label{xdx}
x\frac{d}{dx}\,,\quad (ax+bx^{2})\frac{d^{2}}{dx^{2}}
\ee
These operators play an important role in solving PDEs.

The computation of the derivative of a function $f$ obtained by a
matrix multiplication needs a number of multiplication $\propto
N^{2}$. Actually, the particular structure of the matrix (\ref{e:mat(d/dx)})
suggests
not to perform the full matrix multiplication but instead to use
the recursion formula
\be \label{e:recur_der}
    \hat{f'}_n = 2(n+1) \hat f_{n+1} + \hat{f'}_{n+2} \qquad n \leq N-2 \ ,
\ee
A similar recursion law exists for the operator $d^{2}/dx^{2}$.  In a
such a way the number of operations becomes $\propto N$.  Hereafter we
shall call {\em fast} an algorithm requiring a  number of elementary
operations $\propto N$ or $\propto N\ln N$ where $N$ is the NDF.

\subsection{Primitive of a function} \label{s:prim}

A primitive $F(x)$ of a function $f(x)$ can be obtained by inversion
of the operator $d/dx$, i.e. by solving the algebraic system of $N$
equations
\be \label{e:sys_primitive}
	\sum_{n=0}^{N-1} D_{mn}b_{n} = a_{m} \ , 
\ee
where $b_{m}$ are the expansion coefficients of $F$. The determinant of
$D_{nm}$ vanishes because the primitive of a function is defined within
some additive constant. We can add an algebraic condition, for instance
$F(-1) = 0$, in order to eliminate this degeneracy. One (certainly not 
unique !) way to proceed is to  seek
for a primitive that has some property in the coefficient space, for
instance that its first coefficient $b_0$ vanishes. That is
equivalent to find a particular solution $F_p$ of the differential
equation $dF/dx = f$.  Once this solution is found, we can add to it a
homogeneous function $F_{h}$ in such a way that the final solution $F =
F_p + F_h$ fulfils the required conditions. In this example $F_{h}$ is
constant. If we look for a particular solution having $b_0 = 0$, the
new system is
\def\arraystretch{1.}\label{Dmn2}
\be
\arraycolsep=3mm
\left[\begin{array}{rrrrrrr}
      1&    0&     0&    0&     0&    0&     0\\
      0&    2&     0&    0&     0&    0&     0\\
      0&    0&     4&    0&     0&    0&     0\\
      0&    0&     0&    6&     0&    0&     0\\
      0&    0&     0&    0&     8&    0&     0\\
      0&    0&     0&    0&     0&   10&     0\\
      0&    0&     0&    0&     0&    0&     12
\end{array}
\right]
\def\arraystretch{1.}
\left[\begin{array}{r}
      b_{0}\\
      b_{1}\\
      b_{2}\\
      b_{3}\\
      b_{4}\\
      b_{5}\\
      b_{6}
\end{array}
\def\arraystretch{1.}
\right]
=\left[\begin{array}{r}
              0\\
   \tilde{a_{0}}\\
   \tilde{a_{1}}\\
   \tilde{a_{2}}\\
   \tilde{a_{3}}\\
   \tilde{a_{4}}\\
   \tilde{a_{5}}
\end{array}
\right]
\ee
where $\tilde{a_{j}}$ is the new RHS vector corresponding to the 
linear combination
\be
	\tilde D_{mn} = D_{mn} - {1\over 1+\delta_m^0} D_{m+2,n}
		\qquad 0\leq m \leq N-4 \ ,
\ee
which transforms the matrix $D_{mn}$ into a diagonal one. 
It follows that the
computation of the primitive of a function can be performed with a fast
algorithm.
This technique is
quite general and is similar to that used when treating more complicated
problems (cf. Sect.~\ref{s:BC_impl}).

\subsection{Reduction to band matrices of more general operators}
\label{s:reduc_oper_gen}

The matrix representation $D^{2}_{mn}$ of the operator $d^2/dx^2$
can also be transformed to a band matrix with some simple linear
combination of the lines. Applying first the linear combination
\be\label{eq:d2a}
\tilde{D}^{2}_{mn}= { (1+\delta_0^m) D^2_{mn} - D^2_{m+2,n} \over m+1} 
	\qquad 0 \leq m \leq N-3 
\ee
followed by
\be\label{eq:d2b}
\bar{D}^{2}_{mn}=\tilde{D}^{2}_{mn}-\tilde{D}^{2}_{m+2\,n}\qquad 0\leq m \leq 
N-5
\ee
transforms the matrix into
\def\arraystretch{.9}
\be
\arraycolsep=3mm
\bar{D}^{2}_{nm}=\left[\begin{array}{rrrrrrrrr} 
      0&     0&     8&     0&     0&     0&     0\\
      0&     0&     0&    12&     0&     0&     0\\
      0&     0&     0&     0&    16&     0&     0\\
      0&     0&     0&     0&     0&    20&     0\\
      0&     0&     0&     0&     0&     0&    24\\
      0&     0&     0&     0&     0&     0&     0\\
      0&     0&     0&     0&     0&     0&     0\\
\end{array}
\right]
\ee
More generally, the matrix representation of the differential operator
\be\label{genoper}
{\bf L}={\bf I} +(a_{0}+a_{1}x)\frac{d}{dx}+(b_{0}
	+b_{1}x+b_{2}x^{2})\frac{d^{2}}{dx^{2}}  \ , 
\ee
where ${\bf I}$ is the identity and $a_0$, $a_1$, $b_0$, $b_1$ and $b_2$ 
are numerical
constants, can be transformed to a penta-diagonal matrix via the linear
combinations (\ref{eq:d2a}) and (\ref{eq:d2b}). Therefore there exists
a fast algorithm to inverse the above operator. Note that the operator
${\bf I}$ is responsible for the five diagonals of the matrix.

Finally, the multiplication and division of a function by $X=a+bx$  can
be performed in the coefficient space. The matrix representation of 
this operation 
is a bi-diagonal matrix $X_{ij}$. Therefore multiplication and division
are performed with a fast algorithm. The advantages in operating in
this way are the following ones. Firstly, it is easier to handle the division
by $X$ when $X$ vanishes for a given value of $x$. Moreover, the
roundoff error is less important than when performing the
multiplication in the configuration space. Secondly, computing quantities
like $(x+1)^{l}$, ($l$ integer $\geq 1$) in the configuration space
introduces high frequency terms. In fact, for
large values of $l$, the function $(1+x)^{l}$ is numerically zero and
therefore not continuous near $x = -1$. Such functions play an important
role in spherical coordinates, because  they are the radial part of the
spherical harmonics of order $l$. Thirdly, it avoids to come back to the
configuration space to perform the multiplication by $x$ when the
function is known in the coefficient space.

\section{Solution of elementary P.D.E. in Cartesian coordinates}
\label{s:solution_PDE}

\subsection{Temporal scheme}\label{s:Tempor}

Consider the 1-D second  order hyperbolic PDE (heat equation with
advection)
\be\label{e:heat}
\frac{\partial\Theta}{\partial t} + v(x)\frac{\partial \Theta}{\partial x}
-\mu(x)\frac{\partial^{2}\Theta}{\partial x^{2}}=S(x,t) 
	\quad x \in [-1,+1],\ t \ge 0 \ . 
\ee
If $\mu(x)=0$ and $v(x)=v_{0}={\rm const}$, the general solution is
$\Theta(x,t)=f(x-v_{0} t)$ i.e. a wave propagating from the left to the
right if $v_{0} > 0$ and from the right to the left in the opposite
case. The time evolution of $\Theta$ can be obtained by computing the
time derivative with a finite difference method: $t$ is discretized at
given uniformly spaced instants $t_{j}$, $t_{j+1}=t_{j}+\Delta t$ and
a second order scheme (for example an explicit Crank-Nicolson scheme)
approximates, Eq.~(\ref{e:heat}) by
\be\label{e:Cran}
\Theta^{j+1}-\Theta^{j} = - {\Delta t \over 2}
	\left( v \frac{\partial \Theta } {\partial x}^{j+1} 
	+ v \frac{\partial \Theta^{j}}{\partial x}
	+ S^{j+1/2}
	\right) \ , 
\ee
where $\partial \Theta^{j+1}/\partial x$ at the RHS is computed by
extrapolating its values from the past. This scheme is highly unstable,
no matter the choice of the time-step $\Delta t$. This
instability is due to the absence of boundary conditions. The value of
$\Theta$ at $x=-1$ if $v_{0} > 0$ or at $x=1$ in the opposite case, must
be given. In other words, we have to define what goes into the interval
$[-1,+1]$.

Before to explain which boundary condition makes the problem well posed
(in the sense given by Hadamard, see e.g. \cite{Pet54} p. 62) and how
it can be implemented, it is worth to discuss the Courant conditions
that garanties the numerical stability.

In the case of Chebyshev polynomial expansion, the numerical stability
is achieved if $\Delta t \propto 1/N^{2}$. This condition (Courant
condition) is much more severe than in the case of finite difference
method with uniform grid for which $\Delta t \propto 1/N$. The
situation is much worse if a spatial second order operator is present
($\mu(x) \ne 0$). In this case $\Delta t$ must decrease as $1/N^{4}$.
These severe Courant condition is due to the fact that the density of
the Chebyshev collocation points behave like $1/N^2$ near the
boundaries $x=-1$ and $x=+1$ (see e.g. Fig.~\ref{f:u,I4,I8} or 
Fig.~\ref{f:advs1}). However, if the coefficients of the space
derivatives vanish at the boundaries, the Courant condition becomes
$\Delta t \propto 1/N$ in the case of a first order spatial
differential operator and $\Delta t \propto 1/N^{2}$ in the case of a
second order spatial differential operator. Note that 
the dense sampling at the boundaries can also be an advantage: it
is well suited to
resolve boundary layers if they are present in the solution.  We shall
give an example below (Sect.~\ref{s:Layer}). 

Implicit (or semi-implicit) schemes allow to overcome this difficulty.
Implicitation of the Crank-Nicolson  scheme (\ref{e:Cran}) can be
obtained by inverting the operator $Id+\Delta t \partial/\partial x$.
The generalization to the case $\mu(x)=\mu_{0} =const. \ne 0$ is
straightforward.  With an implicit scheme the absolute convergence (for
any value of $\Delta t$) is obtained.

The leapfrog scheme, which is very widely used with Fourier expansion,
turns out to be unstable with Chebyshev expansion. The reader will find
an exhaustive discussion on this subject in \cite{GotO77} pp. 103-115.
In the present article, we shall use systematically the Crank-Nicolson
scheme. Its accuracy behaves like $\Delta t^{3}$ per time-step when all
the quantities are computed at the time $t^{j+1/2}=t^{j}+\Delta t/2$.

In a more general case, when the coefficients of the spatial
derivatives depend on $x$, the implicitation consists to invert an
operator of the kind ${\cal O} = Id+b(x,t) D$ where $Id$ and $D$ are
respectively the identity and a differential operator and where
$b(x,t)$ is some function. The matrix representation of operator $\cal
O$ is in general a full matrix which cannot be reduced to a band matrix
by means of simple operations. Consequently, the inversion of $\cal O$
operator requires a number of operations $\propto N^{3}$. In order to
save CPU time and to build a fast algorithm, the following
{\em semi-implicit} method is highly recommended.

Consider for simplicity Eq. (\ref{e:heat}) with $v(x)=0$. If the
Crank-Nicolson scheme is used, the corresponding PDE reads
\be\label{e:simpa}
\Theta^{j+1}-\frac{1}{2}\Delta t\,\mu(x)\frac{\partial^{2}\Theta^{j+1}}
{\partial x^{2}}
	= \Theta^{j} + \frac{1}{2}\Delta t\,\mu(x)\frac{\partial^{2}\Theta^{j}}
	{\partial x^{2}}+S(x)^{j+1/2} \ .
\ee
The above time discretization can be reformulated in the equivalent form 
\begin{eqnarray}
\Theta^{j+1}-\frac{1}{2}\Delta t\, \mu^{max}\frac{\partial^{2}\Theta^{j+1}}
{\partial x^{2}}
	& = & \Theta^{j} + 
		\frac{1}{2}\Delta t\, \mu(x)\frac{\partial^{2}\Theta^{j}}
		{\partial x^{2}} \nonumber \\
	&& -\frac{1}{2}\Delta t(\mu^{max}-\mu(x))\frac{\partial^{2}
		\Theta^{j+1}}{\partial x^{2}}+
		S(x)^{j+1/2} 	\ , 			\label{e:simpb}
\end{eqnarray} 
where $\mu^{max}$ is the maximum value of $\mu(x)$ on the interval
$[-1,+1]$. The RHS of Eq. (\ref{e:simpb}) becomes the source of the new
equation. The term $(\mu^{max}-\mu (x))\partial^{2}\Theta^{j+1}
/\partial x^{2}$ is computed by means of a second order extrapolation
from its value at the items $t^{j-2}, t^{j-1},t^{j}$. In this way,
we reduce the problem to the case of constant coefficients.

It is obvious that the new temporal scheme is of second order too.
However, its precision depends on how close $\mu(x)$ is to
$\mu^{max}$. Moreover, this method cannot be applied straightforwardly to
the case where $\mu(x)$ vanishes at the boundary of the interval.

To improve the method, we can introduce a polynomial $P_2(x)=a_{0} +
a_{1}x+a_{2}x^{2}$ instead of $\mu^{max}$ which behaves as $\mu(x)$ at
the boundaries and such that $P_{2}(x)-\mu(x)\ge 0$. As mentioned in
Sect.~\ref{s:reduc_oper_gen}, the matrix representation of the operator
$Id+P_{2}(x)\partial^{2}/\partial x{^2}$ can be reduced to a
penta-diagonal matrix by means of the linear combination
(\ref{eq:d2a})-(\ref{eq:d2b}). The generalization to the case $v(x)\ne 0$
is obvious.

Implicit or semi-implicit temporal scheme in multi-dimensional problems
can be easily implemented with the Alternating Direction Method (cf.
\cite{RicM67}). We have tested this method
for 2-D problems and did not found any difficulty. We conjecture
that this method can be used for 3-D problems too.

\subsection{Boundary conditions and well posed problems}\label{s:BC}

Because of its generality, we will use equation (\ref{e:heat}) to
discuss in detail the numerical problems tied to boundary conditions
and its solution.

The problem is well posed in the following cases.  

\begin{itemize}

\item If $\mu(x)=0$, the equation reduces to a first order PDE. In
this case, if $v(x)\ne 0$ for $x \in [-1,+1]$ then one BC at $x=-1$ (
$v(x) > 0$) or at $x=+1$ ($v(x) < 0$) must be imposed. \\[0mm]

\item If $\mu(x)=0$ and 
	$v(x)\geq 0$ for $x \in [-1,+1]$ and $v(-1)=0$ then no BC can be
imposed. The same  holds if $v(x) \leq0$ for $x \in [-1,+1]$ and $v(+1)=0$.
\\[0mm]

\item If $\mu(-1)=\mu(+1)=v(x)=0$ but $\mu(x) > 0$ for $x \in ]-1,+1[$,
the rules given for the first order PDE hold.\\[0mm]

\item If $\mu(x)$ and $v(x)$ vanish on the boundaries, only one BC
is required (as in the previous case).\\[0mm]

\item If $\mu(x) > 0$ everywhere, two BC are required.

\end{itemize}

For a second order PDE, boundary conditions are quite general and can
have the form
\be\label{e:B.C.}
	\alpha (t)\frac{\partial \Theta}{\partial x} + 
			\beta(t)\Theta = \gamma(t) \ , 
\ee
but other non local BC (dual BC) can be chosen (see below).  An example
of a not well posed problem is to impose two BC (the value of $\Theta$
and its derivative) at one boundary of the interval.

\subsection{Boundary layer and dual boundary conditions}\label{s:Bol}

Consider Eq. (\ref{e:heat}) with $\mu$ and $v$ constant:
$\mu(x)=\mu_{0}$ , $v(x)=v_{0} >0$. If $\mu_{0} \to 0$, the equation
degenerates in a first order equation. We can not impose two BC in the
case of a first order equation. Physically, when $\mu_{0} \to 0$, a
boundary layer is formed at $x=+1$ and the numerical algorithm has been
able to describe the strong variation of the solution near the
boundary. We can form an adimensional parameter $N_{p}$ 
(the {\em P\'{e}clet number}) which characterizes the thickness of the layer~:
$N_{p}=v_{0}L/\mu_{0}$ where $L$ is the typical dimension of the
system. The thickness $D_{l}$ of the layer is $D_{l}=L/N_{p}$. In this
example, $L=2$ and $N_{p} =2v_{0}/\mu_{0}$. Thanks to the accumulation
of collocation points at the boundary, Chebyshev expansion is well
suited to treat this problem. The layer is well resolved and the
maximum value of $N_{p}$ which garanties numerical stability scales as
$N^{7/4}$. For instance with $N=33$ collocation points $N_{p}$ can be
as high as $200$ (see e.g. \cite{GotO77} p. 140).

In physical applications, it can happen that a fine resolution of the
boundary layer is not required. The natural question arises how modify
the BC in such a way that the new solution differs appreciably from the
exact one only in the layer. It is clear that the boundary layer is
created by the BC that not would exist in the case $\mu_{0}=0$.
Unfortunately, numerical instabilities appear if only one BC is
imposed when $\mu_{0} \ne 0$. To overcome this difficulty, we need to
find a criterium which leads to a value of $\Theta$ at the boundary
where the layer forms ($x=+1)$ such that the numerical scheme is stable
if $\mu_{0}=0$ too. In other words, we need a redundant BC.

Such a BC is more general than the usual one. We will call it a ``non
local'' boundary condition because this condition is imposed in the
coefficients space. There is a wide class of criteria to find a magic
value for $\Theta(+1)$. One criterium consists in finding $\Theta(+1)$
such that, at each time step, $\Theta(x)$ is as smooth as possible. For
instance, one can minimize the norm of $\partial^{2}\Theta(x)/\partial
x^{2}$ i.e.
\be
\int_{-1}^{1}\left(\frac{\partial^{2}\Theta(x)}{\partial
x^{2}}\right)^{2}dx \ , 
\ee

Another method  consists to cancel the last coefficient of the
expansion of $\Theta$ or to minimize the sum of the square of the last
$J$ coefficients. All these exotic BC garanties the stability of the
numerical scheme for any value of $\mu_{0} \ge 0$ and gives very
similar results. Note that the last two criteria introduce an
evanescent error when $\mu(x)=0$.

An heuristic justification of this procedure can be easily understood
by noting that the P\'{e}clet number parametrizes the antagonism
between the advection term $v_{0} > 0$, which carries the information
from the left to the right, and the diffusion term $\mu_{0}$ which
diffuses back the information. For high P\'{e}clet numbers, the
diffusion term influences only the boundary layer. Therefore, the
solution outside of the layer is almost not influenced by the BC.

The above exotic BC are useful in other physical situations.
Consider a region $R$ of the interior of a star. Let us suppose that we
are interested in phenomena described by a PDE of the kind of Eq.
(\ref{e:heat}) only in this region. We need to know which BC to impose
at the boundary of $R$. This BC cannot be known without solving the
equation in {\em all} the star. We know only that the solution is
smooth in the interior of the star. Therefore, we can impose a value of
the solution which makes the solution as smooth as possible. Roughly
speaking, the underlying philosophy is to introduce the minimum of
information when the problem is not completely determined. The reader
will recognize the same philosophy which underlies the maximum entropy
criterium used in the theory of signal.

\subsection{Implementation of the boundary conditions}\label{IBC}

There are various ways to implement the BC within spectral methods. 
We present here three different methods. 

\subsubsection{Galerkin method}

The {\em Galerkin method} consists in choosing the set of functions used for
the expansion a set of function satisfying the required BC. The example
showed in Sect.~\ref{s:fexp} solves a PDE with  periodical BC by
means of Fourier expansion. The Galerkin method will be widely used to
handle regularity conditions (e.g. with spherical coordinates where
coordinate singularities appear). It is sometimes easy to find a set
of functions satisfying the desired BC by taking a linear combination
of Chebyshev polynomials. For instance, if the solution of the PDE must
vanish at $x=-1$, one can expand the quantities onto the new basis of
function $\Phi_{n}$~:
\be\label{eq:gal}
\Phi_{n}(x)=T_{n}(x)+T_{n+1}(x) \ . 
\ee
Note that the  matrix of the corresponding linear algebraic system of
equations can be reduced to a band matrix via the linear combinations
(\ref {eq:d2a}),(\ref{eq:d2b}). However, the number of diagonal is
larger: the $5$ diagonals for Eq.(\ref{e:heat}) become 7 when the above
Galerkin basis is used.

\subsubsection{Lanczos method}

The {\em Lanczos method} \cite{Lan56} consists to replace the source
$S_{n}^{j+1/2}$ (in the coefficient space) of Eq. (\ref{e:heat}) by
\be\label {e:Lanc}
{\tilde S}_{n}=S_{n}^{j+1/2}+b_{1}\delta_{N-1,n}+b_{2}\delta_{N,n} \ ,
\ee
where $\delta_{n,m}$ is the Kronecker symbol and to compute, at
each time step, the coefficients $b_{1}$ and $b_{2}$ in such a way that
BC are fulfilled (cf. Sect.~\ref{s:BC_impl}). 

\subsubsection{Dual Lanczos method}

A {\em dual Lanczos method}\footnote{In the book by Gottlieb \& Orszag
\cite{GotO77}, this method is called {\em collocation approximation}
(page~14 of \cite{GotO77}). We prefer the term {\em dual Lanczos method}
for obvious reasons.} consists to add in the configuration space two Dirac's
functions to the solution. The coefficients of these $\delta$ function
are computed in order to satisfy the BC. The source $S^{j+1/2}(x_{k})$
is replaced by
\be\label{e:Land}
{\tilde S}(x_{k})=S(x_{k})^{j+1/2}+b_{1}\delta(x_{k},-1)
		+b_{2}\delta(x_{k},1)
	\ . 
\ee 

These three methods can be used either with explicit or implicit
temporal scheme. From the numerical point of view, the implicit scheme
must be treated carefully, especially for second order equations when the
time step is large.

\subsection{Details about the boundary condition implementation}\label{s:BC_impl}

Let us consider the heat equation (\ref{e:heat}) with $v(x)=0$ and $\mu(x)=1$
and let us suppose that the time discretization is performed according to a 
first order fully implicit time scheme. For $\Delta t = 0.1$, the system
to be solved is then (cf. Eq.~(\ref{e:Mat2})):
\be\label{s:lay}
\arraycolsep=2.25mm
\def\arraystretch{1.05}
\left[\begin{array}{rrrrrrr}
      1&    0&   -.4&    0&  -3.2&    0& -10.8\\
      0&    1&     0& -2.4&     0&  -12&     0\\
      0&    0&     1&    0&  -4.8&    0& - 19.2\\
      0&    0&     0&    1&     0&   -8&     0\\
      0&    0&     0&    0&     1&    0&   - 12\\
      0&    0&     0&    0&     0&    1&     0\\
      0&    0&     0&    0&     0&    0&     1
\end{array}
\right]
\def\arraystretch{1.}
\left[\begin{array}{r}
      \Theta_{0}\\
      \Theta_{1}\\
      \Theta_{2}\\
      \Theta_{3}\\
      \Theta_{4}\\
      \Theta_{5}\\
      \Theta_{6}
\end{array}
\def\arraystretch{1.}
\right]
=\left[\begin{array}{r}
    \tilde{S_{0}}\\
   \tilde{S_{1}}\\
   \tilde{S_{2}}\\
   \tilde{S_{3}}\\
   \tilde{S_{4}}\\
   \tilde{S_{5}}\\
   \tilde{S_{6}}
\end{array}
\right] \ , 
\ee
where  the R.H.S. of the system, $\tilde{S}_{n}$, 
is the Lanczos modified source 
defined by Eq.~(\ref{e:Lanc}) or by Eq.~(\ref{e:Land}) depending on
the choice of the BC treatment method.
The simplest way to find the value of the coefficients $b_{1}$ and $b_{2}$
is to solve  first the system having as source the vector $S_{n}$ only
($b_{1}=b_{2}=0$). In this way one particular solution $\Theta^{\rm par}_{n}$ 
of the  system is found. The second step consists in finding two homogeneous
solutions $\Theta^{\rm h1}_{n}$ and  $\Theta^{\rm h2}_{n}$ by solving twice 
the system (\ref{s:lay}) with the source $\delta_{N-1,n}$ and $\delta_{N-2,n}$ 
(for the Lanczos 
method) and finally by computing $b_{1}$ and ${b_2}$ in order to 
fulfil the required BC. 
The same procedure holds for the dual Lanczos method. We recall that
the Chebyshev coefficients of the Dirac functions at the
interval boundaries, namely $\delta(x+1)$ and $\delta(x-1)$, are
respectively $\hat \delta_{n}=(-1)^{n-1}$ and $\hat \delta_{n}=1$.
Of course, as already said, the matrix of the above system can be reduced 
with  simple linear combinations to a penta-diagonal matrix and consequently
finding solutions of a banded  system having 3 different R.H.S. is a very 
little time consuming.
The homogeneous solution can be used to satisfy the dual BC by imposing, for
example, that the last coefficient of the solution vanishes.
By looking at the matrix of the system (\ref{s:lay}), it appears that the 
extra-diagonal coefficients for $\Delta t=.1$ are much larger than the 
coefficients of the main diagonal. It follows that each 
solution of the system depends strongly on the value of the highest order
coefficients of the source (this is a consequence of the fact that the
problem is not well posed without BC). Moreover, small values of the highest
order coefficients lead to very large solutions. When  the linear
combination of the 3 solutions required  to  fulfil the correct BC is 
performed compensations occur to give the correct solution. These compensations
are source of important round-off errors. In practice, $\Delta t \propto 
1/N^{3}$ is a necessary condition to have high precision results.
Another procedure works in the opposite case ($\Delta t$ high)
and is easy to implement for the Lanczos method. It is the following one:
look for a solution of the first equation having some condition in the
coefficient space, for example having the two first coefficients vanishing.
This can be obtained easily by replacing the system (\ref{s:lay}) by 
\be\label{s:layb}
\arraycolsep=2.25mm
\def\arraystretch{1.05}
\left[\begin{array}{rrrrrrr}
      1&    0&     0&    0&     0&    0&     0\\
      0&    1&     0&    0&     0&    0&     0\\ 
      1&    0&   -.4&    0&  -3.2&    0& -10.8\\
      0&    1&     0& -2.4&     0&  -12&     0\\
      0&    0&     1&    0&  -4.8&    0& - 19.2\\
      0&    0&     0&    1&     0&   -8&     0\\
      0&    0&     0&    0&     1&    0&   - 12\\
\end{array}
\right]
\def\arraystretch{1.01}
\left[\begin{array}{r}
      \Theta_{0}\\
      \Theta_{1}\\
      \Theta_{2}\\
      \Theta_{3}\\
      \Theta_{4}\\
      \Theta_{5}\\
      \Theta_{6}
\end{array}
\def\arraystretch{1.}
\right]
=\left[\begin{array}{r}
   0           \\
   0           \\
   S_{0}\\
   S_{1}\\
   S_{2}\\
   S_{3}\\
   S_{4}
\end{array}
\right] \ . 
\ee
In this way we have found a particular solution having the two first 
coefficients vanishing.
Find then two other homogeneous solutions having as source $\delta _{1,n}$
and $\delta_{2,n}$  and make a linear combination in order to fulfil
the BC. It is obvious that the solution obtained in this way is the same
than that obtained by solving the original system (\ref{s:lay}):
both solutions satisfy indeed the first $N-2$ equations and the BC; 
the unicity theorem garanties that the two solutions are identical.    

The matrix of the new system has the coefficients of main diagonal which 
increase
when $N$ increases and the solution is computed with high accuracy for 
$\Delta t > 1/N^{4}$. In fact, in the opposite case,  the determinant 
of the new matrix vanishes
for $\Delta t=0$ and the round-off errors become important for too small
values of $\Delta t$. Therefore the solution method depends on the value of
$N$ and $\Delta t$. We recommend the Lanczos method for large values
of $\Delta t$ and the dual method in the opposite case. For first order 
equations, or for the advective diffusive equation with high P\'eclet numbers, 
the dual Lanczos method is more convenient.   

\subsection{Artificial spectral diffusivity}\label{s:artvis}

In classical hydrodynamics with finite difference methods, artificial
viscosity or diffusivity  is often used in order to avoid singularities
in the numerical solutions and/or to stabilize the numerical scheme. 

Within Fourier expansion, a high order operator is added to the RHS of
the PDE in order to obtain a smoother solution. In the Fourier space,
this artificial diffusivity is of the kind $\propto k^{2L}$
\cite{BasLSB81} ($L$ integer number) where $k$ is the spatial frequency.
For $L=1$, this term represents the ordinary diffusivity.  
For $L\to\infty$, its tends to the {\em evanescent viscosity} introduced
by Maday et al. \cite{MadOT93}. Within
Chebyshev expansion, spectral viscosity can be implemented via a power
of the degenerated operator ${\cal V}= \sqrt{1-x^2}d/dx
\left[\sqrt{1-x^2}d/dx\right]$ \cite{MarB92}. This is an extension to 
Chebyshev SM of the technique developed in \cite{BasLSB81} for Fourier SM. 
Its matrix representation is a diagonal matrix:  
${V}_{nm}= - n^{2}\delta_{nm}$. 
Analogous {\em linear} artificial viscosity methods were used in 
numerical magneto-hydrodynamics for handling
sharp gradients induced by non-linearities \cite{HarS86,SchBMHC87,LerL91}.
Non-linear artificial viscosity ({\em hyperviscosity}) has been introduced
in \cite{PasP88} and applied to MHD problems \cite{PasPPS90}.

\section{Examples of time evolution}\label{s:Exa}

\subsection{Advection equation}\label{s:adve}

\begin{figure}
\centerline{\epsfig{figure=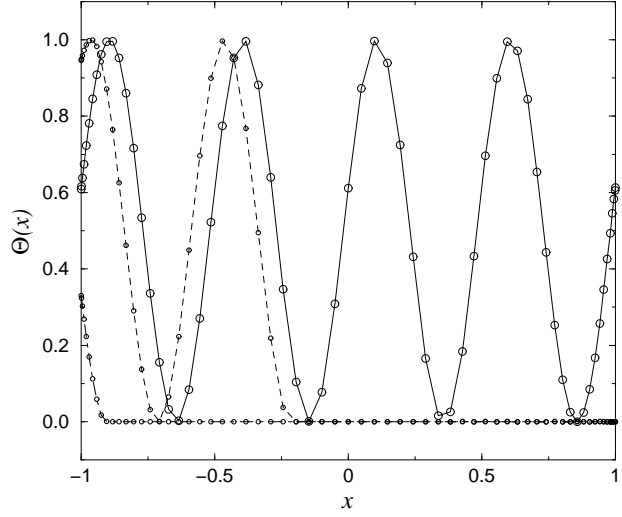,height=8cm}}
\caption{\label{f:advs1} Profiles at different time steps (long dashed, 
short dashed and solid line successively) of the solution
of the advection equation (Eq.~(\ref{e:heat}) with $v(x)=1$ and
$\mu=0$). The initial conditions are $\Theta(x,0)=0$ and the BC is
$\Theta(-1,t)=\sin(4\pi t)^{2}$.  A sinusoidal wave propagates from the
left to the right of the interval.  The collocation points are shown on
the plots ($N=65$, $\Delta t=5\times 10^{-3})$. }
\end{figure}

\begin{figure}
\centerline{\epsfig{figure=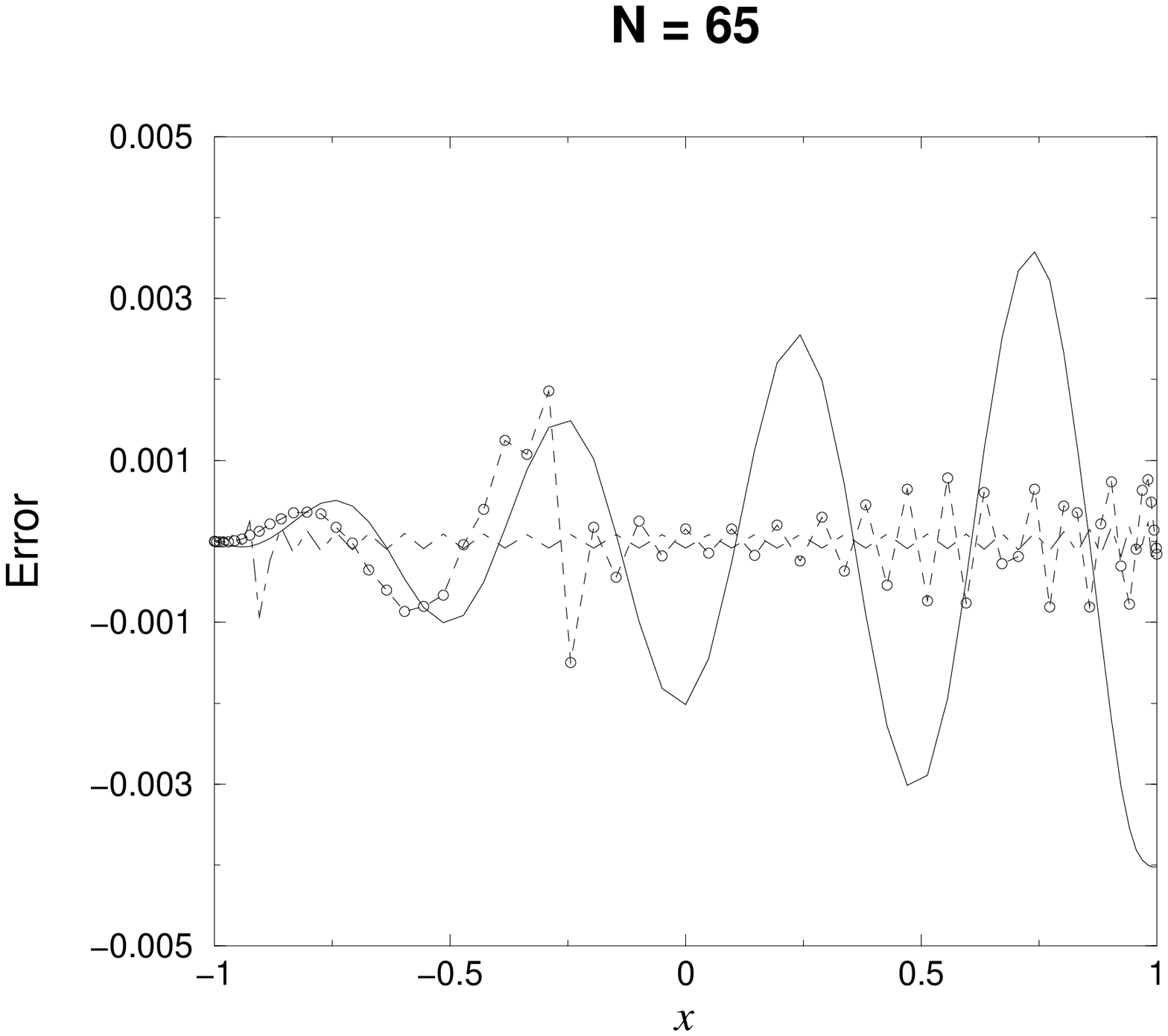,height=8cm}}
\caption{\label{f:adve1} Profiles at different time steps of the error of 
the numerical solution shown in Fig.~\ref{f:advs1}. 
A Gibbs phenomenon appears when the wave is not arrived at the end of interval
yet (see text).}
\end{figure}

In view to compare the different approximations we start with the
advection equation (\ref{e:heat}) with $\mu(x)=0$ and $v(x)=1$. We
choose $\Theta (x,0)=0$ and $\Theta (-1,t)=\sin(4\pi t)^{2}$. The
analytical solution of this PDE is $\Theta_{\rm ana}(x,t)=\sin(4\pi
(t-x))^{2}$ and will be used to compute the numerical error. Spectral
methods are not well suited to solve hyperbolic equations; this example
is worth to show why. Figure~\ref{f:advs1} shows the solution computed
by means of a Chebyshev expansion and an implicit Crank-Nicolson scheme, 
with the Lanczos approximation (Sect.~\ref{IBC}). The corresponding numerical
error is shown in Fig.~\ref{f:adve1}. The solution is ${\cal
C}^{1}$. Therefore the errors decreases as $1/N^{2}$ and Gibbs
phenomenon is heavily present. The solution becomes $C^{\infty}$ only
when the front of the wave has crossed the right boundary.
In the present case, $N=65$ and $\Delta t=5\times 10^{-3}$.

When the solution is ${\cal C}^\infty$, 
the error is due to the temporal scheme which is a second order
implicit one. Decreasing $\Delta t$ by a factor $2$ reduces the error by
a factor 4. From Fig. \ref{f:adve1}, it appears that the error
increases linearly as the wave propagates. This is due to the implicit
scheme which introduces a dissipative term, which gives an
error on the amplitude of the wave. An explicit
leap-frog scheme would avoid this drawback. However, whereas
it is possible to
use a leapfrog temporal scheme for Fourier expansion, such a scheme
is unconditionally unstable for Chebyshev expansion  (see \cite{GotO77} p.
109). Note that with $N=21$, the error due to the computation of the
spatial derivative is still less than the one due to the temporal
scheme when the solution becomes ${\cal C}^{\infty}$.

We conclude that
implicit methods should be avoided for solving hyperbolic equations
when the phase must be computed with a high accuracy. Finally, it is
interesting to compare the above results with the one obtained with
finite difference methods. 
Such a comparison can be found in \cite{GotO77} p.~135. Unfortunately,
these authors compare the results obtained with SM versus 
that obtained by finite difference methods by means
of the ${\cal L}_{2}$ error and, therefore, do not give the error on
the phase and amplitude of the wave.

In the second example, we compare the solutions obtained with the the
Lanczos and dual Lanczos approximations and study the effect of the
spectral viscosity. Let us consider again the advection equation with
same initial condition $\Theta(x,0)=0$ 
but with a sudden excitation at $x=-1$. The
solution is a Heaviside function traveling from the
left to the right.

\begin{figure}
\centerline{\epsfig{figure=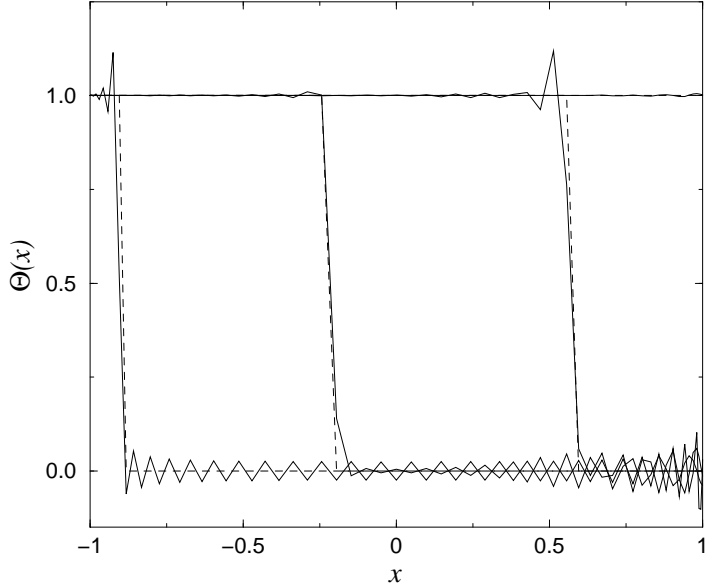,height=8cm}}
\caption{\label{f:ech1} Profile at different time steps of the Heaviside
wave propagating from the left to the right and computed by means of Lanczos 
approximation. The solid line is the numerical solution and the dashed line 
the exact one.}
\end{figure}

\begin{figure}
\centerline{\epsfig{figure=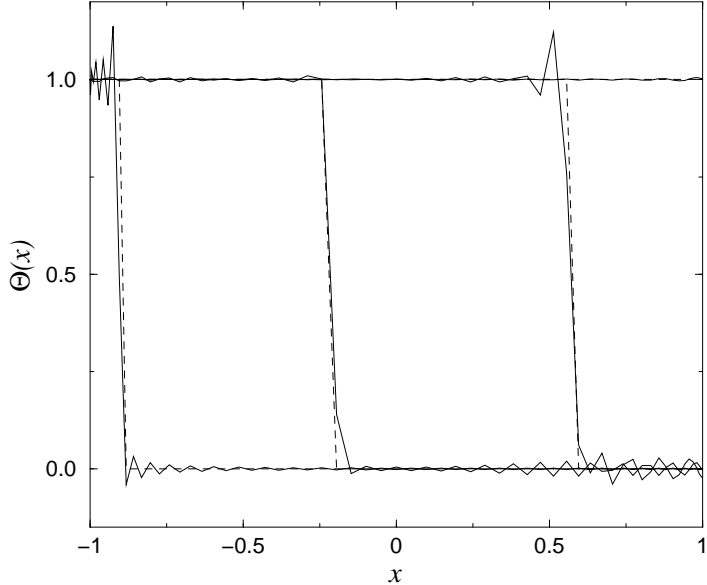,height=8cm}}
\caption{\label{f:ech2} Same as in Fig. \ref{f:ech1} but with the dual 
Lanczos approximation.}
\end{figure}

\begin{figure}
\centerline{\epsfig{figure=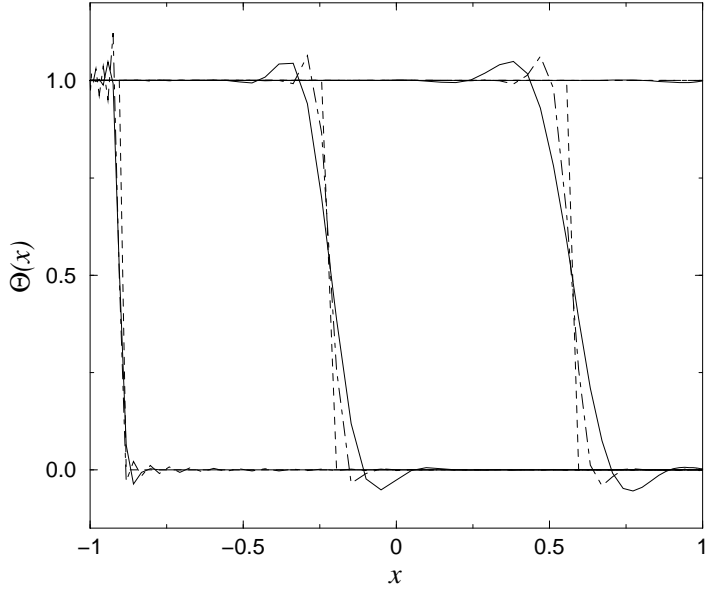,height=8cm}}
\caption{\label{f:ech3} Same as in Fig. \ref{f:ech1} but with dual Lanczos 
approximation and spectral viscosity. The two profiles, beside the analytical
solution (dashed line), correspond to two different
values of the spectral viscosity.  
Note that the discontinuity is smeared out. In spite of the
strong spectral diffusivity, the velocity of propagation of the shock is still
correctly computed. }
\end{figure}

Figure~\ref{f:ech1} shows the solution computed with the Lanczos
approximation. Note that, in spite of a strong Gibbs phenomenon, the
propagation velocity is correctly computed. Figure~\ref{f:ech2} shows the
solution computed with the dual Lanczos approximation (Sect.~\ref{IBC}).
The Gibbs
phenomenon is much less important. Figure~\ref{f:ech3} shows the effect of
the spectral viscosity (Sect.~\ref{s:artvis}) 
(we have used a viscosity operator equivalent
to ${\cal V}^{2}$). The apodisation effect of the spectral viscosity is
well visible. Note that, in spite of the spectral viscosity, the
velocity of the front wave is still accurately computed.

\subsection{Boundary layers and non local boundary conditions}\label{s:Layer}

\begin{figure}
\centerline{\epsfig{figure=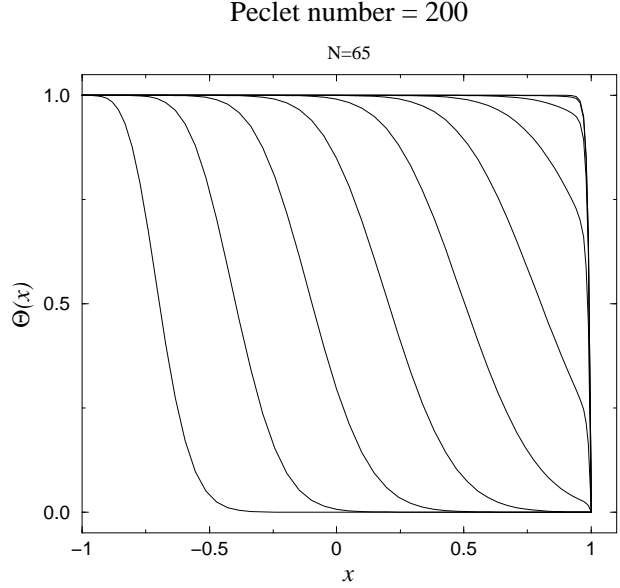,height=8cm}}
\caption{\label{f:heat1} Profiles at different times of the solution of
Eq.~(\ref{e:heat}). Note the formation of the boundary layer near $x=+1$ 
(N=65, $v(x)=1$, $\mu(x)=.01$, $N_{p}=200$, $\Delta t=.015$).}
\end{figure}

\begin{figure}
\centerline{\epsfig{figure=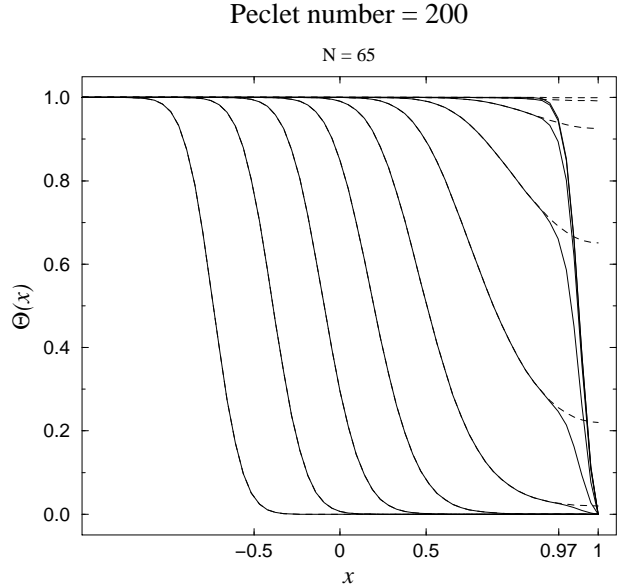,height=8cm}}
\caption{\label{f:heat2} Same as in Fig.~\ref{f:heat1} with different 
boundary conditions. 
The full line is the solution of with two local B.C. and the dashed one is
the solution with a non-local B.C. 
The $x$ axis has been scaled in order to show the structure of the
boundary layer.}
\end{figure}

In this section we give some result obtained by solving the equation
(\ref{e:heat}) in order to show how a boundary layer can be described by
means of SM and to show how non local BC works. Fig. \ref{f:heat1}
shows the solution at different time steps of the equation
(\ref{e:heat}) with $v(x)=1$ and $\mu(x)=.01$. The initial conditions
are the same as in the previous example: $\Theta(x,0)=0$. We suddenly
switch the left BC to the value $1$. The other BC is $\Theta(1,t)=0$. The
Lanczos approximation is used.  The different plots of Fig.
\ref{f:heat1} show the wave propagating from the left to the right. The
solution is similar to the one showed for the advection equation but,
because of the diffusivity, the wave front is less stiff than in the
case of pure advection. When the wave reaches the boundary, the layer
is formed. Its thickness is of the order of $2/N_{p}= 10^{-2}$. 
Figure~\ref{f:heat2} shows the comparison between the solution fulfilling the
above BC (solid line) and that with a non-local BC (dashed line). 
The coordinate $x$ was rescaled in order to show
the structure of the boundary layer. The BC are
$\Theta(-1,t)=1$ and the second one was chosen in order to minimize
the sum of the square of the last three coefficients of the solution.
Note that the new solution differs only inside the boundary layer
($0.97 < x \le 1$). The difference between the two solutions is $\sim
1\times 10^{-12}$ for $x < 0.5$. With a non local BC, the numerical stability
is achieved for any value of $\mu \ge 0$.

\subsection{Kompaneets equation and space compactification}\label{s:Komp}

The last example is the resolution of the Kompaneets equation
(\cite{Mel80}). This equation describes the evolution of the
photon distribution function in a bath of plasma at thermal
equilibrium at temperature $T$ within the Fokker-Planck approximation (see
e.g. \cite{Mel80}). The Kompaneets equation reads:
\be\label{e:Komp}
\frac{\partial {\cal N}}{\partial t}=\frac{\partial}{\partial \nu}\left[
\nu^{2}\beta \frac{\partial {\cal N}}{\partial \nu} -2\beta \nu {\cal N}
 + \nu^{2}{\cal N}
+{\cal N}^{2}\right] \ , 
\ee
where ${\cal N}(\nu,t)$ is the photon density (function of the
frequency $\nu$ ($0\le \nu < \infty$) and the time $t$) and $\beta$ is
the plasma temperature in adimensional units. Its asymptotical solution
is the Bose distribution:
\be\label{e:Bose}
{\cal N}(\nu,\infty)=\frac{\nu^{2}}{\exp(\nu /\beta+\mu)-1} \ , 
\ee
where $\mu \ge 0$ is the chemical potential. It is obvious that the
Kompaneets equation preserves the number of the photons. If the photon
density becomes larger than a critical value, then the photons condense
at the frequency $\nu=0$ (Bose condensation).

To solve the Kompaneets equation, it turns out to be convenient to
compactify the frequency space. This compactification can be performed by
means of a new variable $x=(\nu-1)/(\nu+1)$ ($x\in [-1,1]$,
$\nu=(1+x)/(1-x)$). Equation~(\ref{e:Komp}) then becomes
\begin{eqnarray}\label{Komco}
\frac{\partial {\cal N}}{\partial t}=
  \frac{4\beta \nu^{2}}{(1+\nu)^{4}} \frac{\partial^{2}{\cal N}}
{\partial x^{2}}
&+& \frac{2}{(1+\nu)^{2}}\left[\nu^{2}+2{\cal N}-\frac{2\nu^{2}\beta }
{1+\nu}\right] 
\frac{\partial {\cal N}}{\partial x} \nonumber \\
&+& 2(\nu-\beta){\cal N} \ . 
\end{eqnarray}
If ${\cal N}(-1,0)={\cal N}(1,0)=0$, then 
${\cal N}(-1,t)={\cal N}(1,t)=0$ for all $t\geq 0$ 
and the coefficients of the derivatives vanish
at the boundaries $x=-1$ and $x=1$. 
Therefore, no BC are required. Moreover, the
Courant conditions are fulfilled for $\Delta t\propto 1/ N^{2}$ and
implicitation is not necessary.


\section{3-D decomposition in spherical coordinates} \label{s:3d}

Spectral methods are well suited to handle the pseudo-singularities
tied to spherical-like coordinates $(r,\theta,\phi)$ at
$r=0$ and $\sin\theta = 0$. The
various scalar, vector and tensor fields involved in any physical
problem cannot be any arbitrary function of these coordinates. Any
quantity has to satisfy the so-called {\em regularity conditions} on
the axis $\sin\theta = 0$ and at the origin $r=0$. These regularity
conditions can be taken into account if an appropriated
basis of spectral expansion is chosen. 
Even if {\em regularity conditions} cannot
be called {\em boundary conditions}, this method to handle
pseudo-singularities can be compared to the Galerkin approximation.

As an example, let us consider a scalar function $f(t,r,\theta,\phi )$,
with $r\in [0, 1]$, $\theta \in [0, \pi]$, and $\phi \in [0, 2\pi[$
where $(r,\theta,\phi)$ are the usual spherical coordinates. Assuming
that $f$ is a regular scalar function means, in the context of the
spectral approximation, $f\in {\cal C}^\omega$, that is~:
\be \label{eq:7}
	f=\sum_{i,j,k} a_{ijk} \, x^i y^j z^k \ , 
\ee
where
\be \label{eq:8}
x = r \sin \theta \cos \phi \ ,\ \ 
y = r \sin \theta \sin \phi \ \ {\rm and} \ \ 
z = r \cos \theta \ .
\ee
are the usual Cartesian coordinates. Now, substituting (\ref{eq:8}) into
(\ref{eq:7}) shows that $f$, considered as a function of
$(r,\theta,\phi)$, reads
\be \label{eq:9}
	f(r,\theta,\phi )= \sum_{j,k,m} a'_{jkm} r^{m+2j+k}
				\sin^{m+2j} \theta
				\cos^k \theta \exp(im\phi) \ .
\ee
In other words, this means that $f$ is (obviously) a periodic function
in the $\phi$-direction:
\be \label{e:en-phi}
	f(r,\theta,\phi ) = \sum_{m} a_{m}(r,\theta) \exp(im\phi) \ , 
\ee
where the coefficients $a_{m}(r,\theta)$ satisfy
\be
	a_{m}(r,\theta) = \sin^m \theta \sum_l a_{lm}(r) {P_l}^m(\cos\theta) \ ,
\ee
where ${P_l}^m$ are the associated Legendre functions. The coefficients
$a_{lm}(r)$ satisfy
\be
	a_{lm}(r) = r^l \sum_j a_{jlm} r^{2j} \ .
\ee

A way to handle the regularity conditions on the axis $\sin\theta = 0$
is to expand the angular part of $f$ in a series of spherical harmonics
\be
	f(r,\theta,\phi ) = \sum_{l,m} a_{lm}(r) {Y_l}^m(\theta,\phi) \ . 
\ee
Such an expansion ensures that any pseudo-singularity involving
$\sin\theta = 0$ is automatically handled. Moreover, spherical
harmonics are eigenvectors of the Laplacian operator $\Delta$. This
property allows easy inversion of this operator which appears in a lot
of physical equations (particularly D'Alembertian operator, dissipative
terms of hydrodynamics equations, Poisson equation for
the gravitational field, see Sect.~\ref{s:poisson-scal}).

\begin{figure} 
\centerline{\epsfig{figure=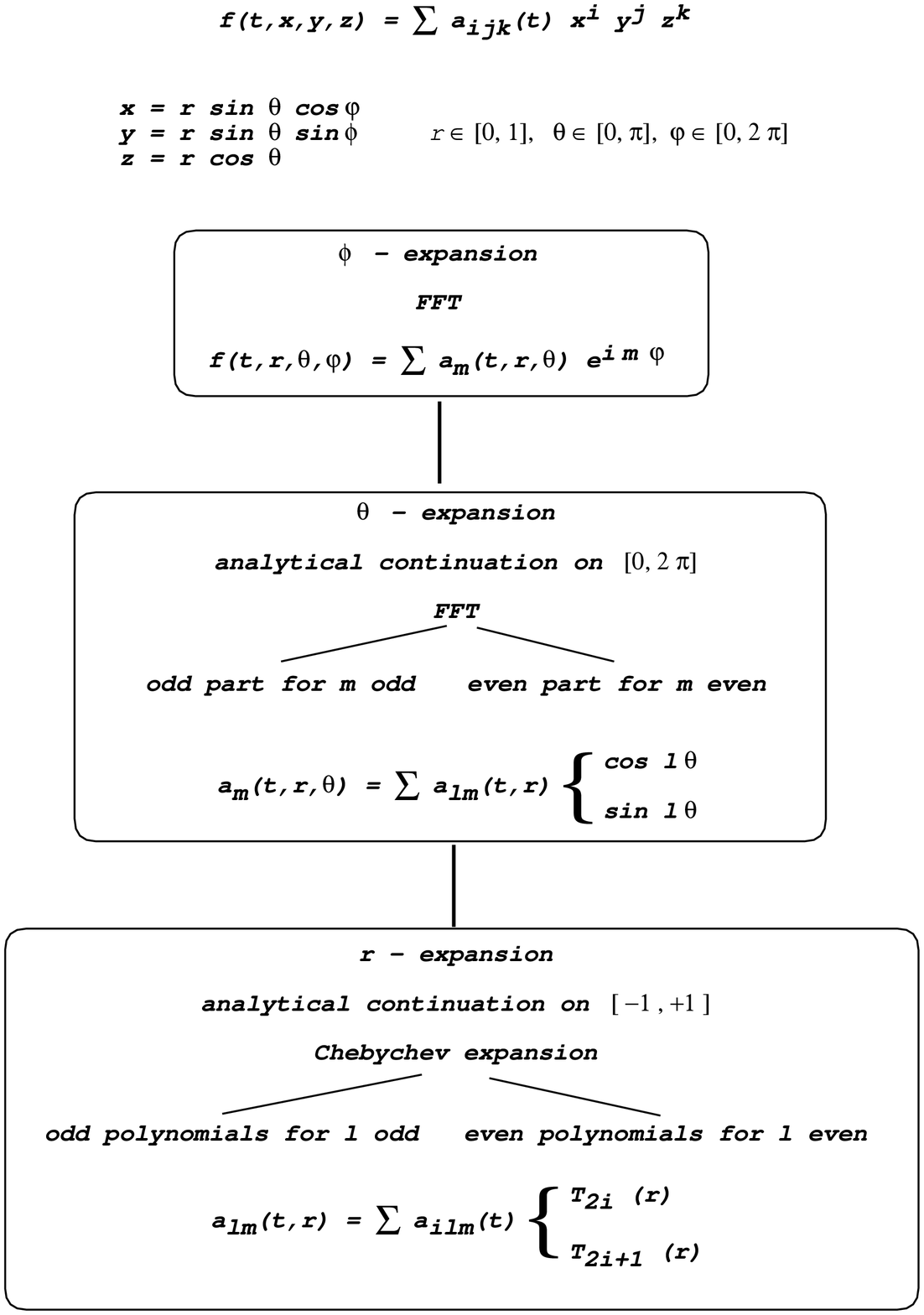,height=12cm}}
\caption{\label{f:deco-3d}
Expansion of a 3-D scalar function in spherical-like coordinates
$(r,\theta,\phi)$.
}
\end{figure}

However, from a numerical point of view, this expansion is not
efficient in the sense that the transformation in the
$\theta$-direction is an expansion in associated Legendre functions
${P_l}^m(\cos\theta)$ which requires $\propto N^2$ operations, $N$ being
the number of degrees of freedom, compared to $N \log N$ operations in
the case of a Fourier transformation (thanks to the FFT algorithm). 

Moreover, it is to be noticed that the regularity conditions described
above, namely that the coefficients $a_m(r,\theta)$ of the expansion
(\ref{e:en-phi}) have to vanish on the axis as $\sin^m\theta$, are not
necessary to handle the singularities present in the differential
operators. More precisely, it is easy to show that any singularity are
automatically handled if the coefficients $a_m(r,\theta)$ satisfy
\be
	a_{m}(r,\theta) = \sin^{{\rm Min}(m,o)}(\theta) 
		\sum_l a_{lm}(r) \cos(l \theta) \ , 
\ee
where $o$ is the order of the differential operator in the
$\theta$-direction. The same considerations hold for the expansion in
the $r$-direction.  

For these reasons, we expand $f$ as a truncated Fourier series in the
$\phi$-direction which requires $\propto N_\phi \log N_\phi$
operations. We perform an analytical continuation of the coefficients
$a_m(r,\theta)$ on the domain $\theta \in [-\pi,+\pi]$ and expand them
in an even (resp. odd) Fourier series in the $\theta$-direction
for even (resp. odd) $m$. This transformation requires $N_\theta
\log N_\theta$ operations. The resulting coefficients $a_{lm}(r)$ are
analytically continued on the domain $r \in [-1,+1]$ and expanded in a
series of even (respectively odd) Chebyshev polynomials depending on the
parity of $l$. This later transformation requires $N_r \log N_r$
operations. The above procedure is detailed in Fig.~\ref{f:deco-3d}. 

\section{Scalar Poisson equation in spherical-like coordinates}              
\label{s:poisson-scal}

We will describe methods which are specifically designed to solve PDE
which contain scalar and vectorial differential operators in a space
domain diffeomorphic to a sphere and in a compactified space. All the
numerical algorithms, but one, are fast algorithms. The only exception
is the Legendre transformation which requires 
$\propto N_r\log N_r N_\theta^{2} N_\phi \log N_\phi$ instead of 
$\propto N_r\log N_r N_\theta \log N_\theta N_\phi \log N_\phi$ operations.

\subsection{Scalar Poisson Equation in a sphere}

Consider the Poisson-like equation in spherical coordinates, 
\be \label{e:poiss}
	\frac{\partial^{2}\Phi}{\partial r^{2}}
	+ \frac{2}{r}\frac{\partial\Phi}{\partial r}
	+ \frac{1}{r^{2}}\left(\frac{\partial^{2} \Phi}{\partial \theta^{2}}
	+ \frac{\cos \theta }{\sin \theta}\frac{\partial \Phi}{\partial \theta}
	+ \frac{1}{\sin \theta^{2}}\frac{\partial^{2}\Phi}{\partial \phi^{2}}
	\right) = S(\vec x,\Phi,\partial \Phi) \ . 
\ee
Solving by iteration this equation by means of spectral methods in a
spherical region is described in \cite{BonGHM92}. Suppose that the source
is known at the $j^{\rm th}$ iteration step. The solution of the Poisson
equation at the $(j+1)^{\rm th}$ step
can be obtained by solving the following system of ordinary
equations on the spherical harmonics coefficients of $S$, say
$S_{lm}(r)$:
\be\label{e:ordif}
\left(\frac{d^{2}}{dr^{2}}+\frac{2}{r}\frac{d}{dr}-\frac{l(l+1)}{r^{2}}\right)
\Phi_{lm}(r)=S_{lm}(r) \ , 
\ee 
where $\Phi_{lm}$ are the spherical harmonics coefficients of $\Phi$.
The involved Legendre expansion is computed from the Chebyshev expansion
in $\theta$ by means of a matrix  multiplication, as detailed in 
Appendix~A.1 of ref.~\cite{BonM90}. 

Galerkin approximation is used to handle the singularity of
Eq.(\ref{e:ordif}) at $r=0$. For $l=0$ the solution is expanded on even
Chebyshev polynomials $T_{2n}(r)$, ($0\le r \le 1$, $0\le n \le N$). 
For $l$ even,
the solution is expanded on the set of functions
$T_{2n}(r)-T_{2n+2}(r)$. These functions as their first derivative
vanish at $r=0$. An analogous treatment is performed for odd values
of $l$ \cite{BonM90}. The matrix representation of the above operator
can be easily reduced to a penta-diagonal one. This matrix has a
vanishing determinant due to the existence of homogeneous solution
$\Phi_{lm}^{\rm h}(r)=r^{l}$.

The BC are satisfied by adding to a particular solution 
a homogeneous solution $\Phi_{ml}^{\rm h}$
which must be computed in the coefficient space
(see Sect.~\ref{s:BC_impl}). Note that when the source does not depend on 
$\phi$ (2-D problems) there exists a fast algorithm for solving the Poisson 
equation that does not need a Legendre expansion (see Appendix A.2 of 
\cite{BonM90}). 
The technical procedure outlined is here is described in details in
Sect.~\ref{s:BC_impl}, since the problem is reduced to a
one-dimensional one for each value of $(l,m)$. 

\subsection{Scalar Poisson Equation in a shell}\label{s:posh}

The solution of the Poisson equation in a shell 
$R_{\rm in}\leq r \leq R_{\rm out}$ is performed in a very
similar way. The main differences are: there is no singularity in the
coefficients of the radial part of Eq.~(\ref{e:ordif}) and there are
two homogeneous solutions $\Phi_{lm}^{\rm h1}(r)=r^{l}$ and
$\Phi_{lm}^{\rm h2}=1/r^{l+1}$. The last analytical  homogeneous solution
can be used to fulfil the two BC. However, it turns out that better
results are obtained by using an homogeneous solution obtained by means
of the Lanczos approximation. In practice an aspect ratio
$R_{\rm out}/R_{\rm in}$ larger than 2 should be avoided.

\subsection{Solution of the Poisson equation in a compactified space}
\label{s:posc}

Some of the Einstein equations lead to Poisson equations for which the source
$S(r,\theta,\phi)$ fills all the space and vanishes at least as
$\propto 1/r^{4}$ when $r \to \infty$. Noting that the solution can be
expanded in power series of the variable $u=1/r$, it is natural to
solve the Poisson equation in two domains $D_{1}$ and $D_{2}$ defined
respectively by $0 \le r \le 1$ and $1 \ge u \ge 0$ and to match the
two solutions and their derivatives at the common boundary. The radial
part of the Poisson operator (with respect to $u$) in the domain
$D_{2}$ reads:
\be
\Delta_{u}=u^{4}\left(\frac{d^2}{du^{2}}-\frac{l(l+1)}{u^{2}}\right) \ . 
\ee
The source must thus be divided by the factor $u^{4}$. This division must be
performed in the coefficient space in order to avoid the singularity at
$u=0$ and to minimize the roundoff errors. The reader will find all
details in \cite{BonGM98a}. Extension of the above technique for more
complicate cases (few domains) are straightforward. Generalization to
parabolic equations is also straightforward. On the contrary, the
above space compactification cannot be used for the wave equation.

\subsection{Poisson equation in a non-spherical domain}
\label{s:non_spher_domain}

Rotating or/and binary stars are not spherical but their surface are
diffeomorphic to a sphere. In order to fulfil the BC and to reduce
truncation errors, it is useful to introduce a system of coordinates
adapted to the geometry of the star. It is easy to find such a
coordinates system, at least for star-like domains (cf. \cite{BonGM98a}).
Let $r=F(\theta ,\phi)$ be the equation of the
surface of the star. Consider the new coordinate system $(\xi,\theta',\phi')$
defined by \cite{BonGM98a}
\be\label{e:transf}
r = a \xi + \xi^{3} \bar{F}(\theta,\phi) + \xi^{4}\hat{F}(\theta,\phi), 
\ \ \theta^{'}=\theta,\ \ \phi^{'}=\phi \ ,
\ee
where $\bar{F}$ and $\hat{F}$ are such that (i) $F(\theta,\phi) = a
+ \bar{F}(\theta,\phi) + \hat{F}(\theta,\phi)$ and (ii) the Fourier 
expansion of $\bar{F}(\theta,\phi)$ (resp. $\hat{F}(\theta,\phi)$) 
with respect to $\phi$ contains only even (resp. odd) harmonics. 
The property (i) ensures that the surface of the star is given by $\xi = 1$,
whereas property (ii) is necessary to have a regular coordinates
transformation. 
Under the coordinates transformation (\ref{e:transf}), the
Poisson equation can be written in the following way:
\be\label{e:poipr}
A\,\Delta^{'}\Phi=B^{ik}\,
\frac{\partial^{2}\Phi}{\partial x^{i}\,\partial x^{k}}+
+C^{i}\,\frac{\partial \Phi}{\partial x_{i}}+S(\xi,\theta',\phi',\Phi,
\partial \Phi) \ , 
\ee
where $\Delta{'}$ is the ordinary Laplacian with respect to the
coordinates $(\xi,\theta'\phi')$ (i.e. $\Delta'$ has the same form as the
operator (\ref{e:poiss}) when $(r,\theta,\phi)$ are replaced by 
$(\xi,\theta'\phi')$) and
$A$, $B$ and $C$ are functions of the coordinates $(\xi,\theta'\phi')$. 
Equation~(\ref{e:poipr}) is solved by means of an iterative
procedure. Since $A$ is not a constant, a procedure similar to that
described in Sect.~\ref{s:Tempor} (Eq.~(\ref{e:simpb})) is used. 
No convergence problems were
found. Note that the above method can be inefficient if the source $S$
does not depend on $\Phi$ and on its derivatives because the iterative
procedure would still be required, whereas the equation would be linear in 
$\Phi$. However, in the
cases of interest, the source $S$ is a non linear function of $\Phi$
and of its first derivatives. Therefore an iterative procedure is
required anyway. The computing time for solving Eq.~(\ref{e:poipr}) is
similar to the one for solving Eq. (\ref{e:poiss}).

\section{Vectorial equations in spherical-like coordinates}\label{s:veceq}

The divergence and curl operators acting on a 3-vector $\vec{V}$ read:
\be \label{e:div}
\vec{\nabla}\cdot\vec{V} =
	\frac{\partial V_{r}}{\partial r}+\frac{1}{r}\left(
	2\,V_{r} +\frac{\partial V_{\theta}}{\partial \theta}+
	\frac{\cos \theta}{\sin \theta}\,V_{\theta}+
	\frac{1}{\sin \theta}\,\frac{\partial V_{\phi}}{\partial \phi} 
	\right)
\ee
and
\begin{eqnarray}\label{e:curl}
\left[\vec{\nabla \land}\vec{V}\right]_{r} &= &\left[
\partial_{\theta}(V_{\phi} \sin \theta)-
\partial_{\phi} V_{\theta}\right]/(r\sin \theta)
\nonumber\\
\left[\vec{\nabla \land}\vec{V}\right]_{\theta} &=& \left[
\partial_{\phi}\,V_{r}-\partial_{r}(r\sin \theta V_{\phi})\right]
/(r\sin \theta)
\nonumber\\
\left[\vec{\nabla \land}\vec{V}\right]_{\phi} &=& \left[
\partial_{r}(rV_{\theta})-\partial_{\theta}\,V_{r} \right]/r \ ,
\end{eqnarray}
where $V_{r},\ V_{\theta},\ V_{\phi}$ denote the components of a
$\vec{V}$ with respect to the orthonormal basis ($\partial_r$,
$1/r\ \partial_\theta$, $1/(r\sin\theta)\ \partial_\phi$) associated with
spherical coordinates $(r,\theta,\phi)$.

It is worth to note a typical behaviour of the spherical components of a
vector. 
Consider a constant vector $\vec{V}$ whose Cartesian components are
$V_{x}=1,\ V_{y}=0, \ V_{z}=0$ . The corresponding spherical components
are $V_{r}=\sin \theta \cos \phi,\ V_{\theta}=\cos \theta \cos
\phi,\ V_{\phi}=-\sin \phi$. From the expression of the divergence
(\ref{e:div}), it is easy to see that there are three singular terms at
$r=0$ and $\theta =0, \pi$ while the sum of these terms is regular.
This is due to the fact that the spherical components of a well
behaved vector are not independent.

A way to overcome this problem is to compute and to add the terms which
generate the singularity before dividing by $r$ and $\sin \theta$ and
to perform the division on the sum. Performing these operations in the
coefficient space allows to resolve the indeterminations at the
singular points and to reduce the roundoff errors. Another method
consists in subtracting to $\vec {V}$ the constant component whose
divergence vanishes. This decomposition can be easily done in the 
coefficient space.

\subsection{Vector decomposition} \label{s:vecdec}

The Clebsch theorem states that a vector $\vec{ V}$ can be uniquely
decomposed into a sum of a divergence-free vector $\vec{W}$ and a
curl-free vector $\vec{ \nabla}\,\Phi$. This decomposition results from
the resolution of a Poisson equation: the divergence of
$\vec{V}=\vec{W}+\vec{\nabla}\,\Phi$ gives indeed $\Delta
\Phi=\vec{\nabla}\cdot \vec{V}$, which has to be solved to get $\Phi$. 
$\vec{W}$ is then computed according to $\vec{W}=\vec{V}-\vec{\nabla}\Phi$. 
The above primitives are computed according to the method exposed in 
Sect.~\ref{s:prim}. 
Once a particular solution
$\vec{W}_{\rm p}$ is obtained, a more general solution satisfying the
desired BC can be obtained by adding to $\vec{W}_{\rm p}$ an homogeneous
solution $\vec{W}_{\rm h}=\vec{\nabla}\,\Phi_{\rm ha}$ were $\Phi_{\rm ha}$ 
is an harmonic function.

\subsection{Solution of the equation $\vec{\nabla}\land\,\vec{V}=\vec{S}$}
	\label{s:eqrot} 

The curl operator $\vec{\nabla}\land$ being degenerated, the vectorial
equation $\vec{\nabla}\land \vec{V}=\vec{S}$ can be solved only if the
divergence of $\vec{S}$ vanishes (integrability condition). Because of
this degeneracy, we can seek for a particular solution $\vec{V}_0$
for which one component vanishes. 

Taking for example $V_{r}=0$, the second equation of the system
(\ref{e:curl}) $\vec{\nabla}\land \vec{V}=\vec{S}$ becomes
$\partial_{r}\,(r\,V_{\phi})=-r\,S_{\theta}$ from which we get
$V_{\phi}=-\frac{1}{r}\,\int_{0}^{r}r'S_{\theta}dr'$. In the same way, we
get $V_{\theta} = \frac{1}{r}\int_{0}^{r}r'S_{\phi}dr'$.  Once this
particular solution is obtained, we can add to it the gradient of some scalar
function in order to obtain a more general solution satisfying the
desired BC.

\subsection{Vectorial Poisson Equation}\label{s:poiv}

The Einstein equations lead to a vectorial Poisson equation for the shift 
vector (Eq.~(\ref{e:eiss2}) below):
\be\label{e:vecp}
\Delta \vec{V}+\lambda \vec{\nabla}(\vec{\nabla}\cdot{V})=\vec{S} \ ,
\ee
where $\lambda$ is some constant. Noting that $\Delta \equiv
-\vec{\nabla} \land\,\vec{\nabla} \land + \vec{\nabla} \, \vec{\nabla}\cdot$ , 
Eq.~(\ref{e:vecp}) can be written as
\be\label{e:vecr}
-\vec{\nabla}\land \,\vec{\nabla}\land \vec{V} +
(\lambda+1)\vec{\nabla}(\vec{\nabla}\cdot\vec{V}) =\vec{S} \ . 
\ee
We shall show how to find the solution of (\ref{e:vecr}) in the
spherical domain ($0\le r \le 1$).  Let us first consider the
degenerated case $\lambda=-1$. The integrability condition is
$\vec{\nabla}\cdot\vec{S}=0$. Let us introduce the vector
$\vec{P}=\vec{\nabla}\land\vec{V}$ which obeys to
$\vec{\nabla}\land\vec{P}=\vec{S}$. Following the method described in
Sect.~\ref{s:eqrot}, we can solve this last equation and get
$\vec{P}=\vec{P}_{\rm p}+\vec{\nabla}(\Psi + \Psi_{\rm ha})$ where
$\vec{P}_{\rm p}$ is a particular solution, $\Psi$ an arbitrary function
and $\Psi_{\rm ha}$ an harmonic function which can be used to fulfil the
BC. The integrability condition imposes that $\Delta
\Psi=-\vec{\nabla}\cdot \vec{P}_{\rm p}$. The final equation for $\vec{V}$
is $\vec{\nabla}\land \vec{V} =\vec{P}+\vec{\nabla}(\Psi+\Psi_{\rm ha})$.
A particular solution $V_{\rm p}$ of this equation (for $\Psi_{\rm ha}=0$) can
be obtained  by means of the method described in Sect.~\ref{s:eqrot}.
The vector $\vec{V}_{\rm g}=\vec{V}_{\rm p}+\vec{\nabla} \Xi$, where $\Xi$ is
some arbitrary function, is again a solution of (\ref{e:vecp}) but is
not the most general one. Actually, we can add to it any solution of
$\vec{\nabla}\land \vec{V}_{\rm ha}=\vec{\nabla}\Psi_{\rm ha}$. 
Such a solution can easily be obtained by means of a spherical harmonics 
expansion of $\Psi_{\rm ha}$.

The case $\lambda \neq -1$ can be reduced to the previous one. Let us
introduce $\vec{V}=\vec{W}+\vec{\nabla} \Phi$ and apply the vector
decomposition as described in section \ref{s:vecdec} to the source.
$\Phi$ can be computed by means of the divergence of both sides of Eq.
(\ref{e:vecp}). Now, the divergence-free counterpart $\vec{W}$ can be
computed as in the previous case ($\lambda = -1$).

\subsection{Telegraph vectorial equation}\label{s:telgr}

The methods described above to solve Poisson equations cannot be
generalized to the parabolic or hyperbolic case. Consider the equation
(telegraph equation)
\be\label{e:telgr}
\alpha\frac{\partial^{2}\vec{V}}{\partial\,t^{2}}+\beta\frac{\partial\vec{V}}
{\partial t}+\lambda\vec{\nabla}(\vec{\nabla}\cdot\vec{V})
+\Delta \vec{V}=\vec{S} \ , 
\ee
where $\alpha$, $\beta$ and $\lambda$ are arbitrary constants. 
After decomposition of the source into a curl-free and a divergence-free 
component, the problem can be split in two parts. This is done mainly
in solving a scalar telegraph equation for the curl-free component and a 
vectorial (with $\lambda=0$)  telegraph equation for the divergence-free
part. 

Eq. (\ref{e:telgr}) explicitely reads:
\begin{eqnarray}
D_{t} V_{r} &+&
\frac{\partial^{2}V_{r}}{\partial r^{2}}
+\frac{4}{r}\frac{\partial V_{r}}{\partial r}+
\frac{1}{r^{2}}\left(2\,V_{r}\frac{\partial^{2} V_{r}}{\partial \theta^{2}}
+\frac{\cos \theta }{\sin \theta}\frac{\partial V_{r}}{\partial \theta}
+\frac{1}{\sin \theta^{2}}\frac{\partial^{2}V_{r}}{\partial \phi^{2}}
\right) \nonumber \\
	&-& \frac{2}{r}\nabla\cdot \vec{V}=S_{r} \label{e:telgc-1} \\
%
%
D_{t}V_{\theta} &+&
\frac{\partial ^{2}\,V_{\theta}}{\partial r^{2}}
+\frac{2}{r}\frac{\partial V_{\theta}}{\partial r}+
\frac{1}{r^{2}}\left(\frac{\partial^{2} V_{\theta}}{\partial \theta^{2}}
+\frac{\cos \theta }{\sin \theta}\,\frac{\partial\,V_{\theta}}{\partial \theta}
+\frac{1}{\sin \theta^{2}}\,\frac{\partial^{2}V_{\theta}}{\partial \phi^{2}}
\right) \nonumber \\
	&+& \frac{1}{r^{2}}\left(
	2\,\frac{\partial\,V_{r}}{\partial \theta}
-2\,\frac{\cos \theta}{\sin \theta}\,\frac{\partial V_{\phi}}{\partial \phi}
-\frac{V_{\theta}}{\sin \theta^{2}}
\right)=S_{\theta} \label{e:telgc-2} \\
%
D_{t} V_{\phi} &+&
\frac{\partial ^{2} V_{\phi}}{\partial r^{2}}
+\frac{2}{r}\frac{\partial V_{\phi}}{\partial r}+
\frac{1}{r^{2}}\left(\frac{\partial^{2} V_{\phi}}{\partial \theta^{2}}
+\frac{\cos \theta }{\sin \theta}\,\frac{\partial\,V_{\phi}}{\partial \theta}
+\frac{1}{\sin \theta^{2}}\,\frac{\partial^{2}V_{\phi}}{\partial \phi^{2}}
\right) \nonumber \\
	&+& \frac{1}{r^{2}\sin \theta}\left
(\cos \theta\, \frac{\partial\,V_{\phi}}
{\partial \theta}+2\,\frac{\partial\,V_{r}}{\partial \phi}
+\frac{2\cos \theta}{\sin \theta}\,\frac{\partial V_{\theta}}{\partial \phi}
-\frac{V{_\phi}}{\sin \theta}\right) = S_{\phi} \ , \label{e:telgc-3}
\end{eqnarray}
where we have introduced $D_{t}\vec{V}\stackrel{def}{=}\,
\alpha\,\partial^{2}\vec{V}/\partial\,t^{2}+ \beta\, \partial\vec{V}/
\partial t$. It easy to see that these equations are badly coupled.
Moreover, there is a lot of singular terms (cf. Sect.~\ref{s:veceq}).
The first equation is coupled with the others by
$\vec{\nabla}\cdot\vec{V}$. Consequently, seeking for a
divergence-free solution to the equation for $V_{r}$ reduces to a
telegraph equation and can be solved {\em mutatis mutandis} by means of
the method described in \cite{BonFG98}. Once $V_{r}$ is known, we can seek
for a solution of the system of coupled equation
governing $V_{\theta},\ V_{\phi}$ whose sources are
\begin{eqnarray}\label{e:angpot}
\hat{S}_{\theta} &=& S_{\theta}-2/r^{2}\,\partial_{\theta}V_{r}
	\nonumber \\
\hat{S}_{\phi}&=& S_{\phi}-2/(r\sin \theta)^{2}\,\partial_{\phi}V_{r} \ . 
\end{eqnarray}

Introducing the couples of potentials $(\Lambda,\Upsilon)$ and $(U,W)$,
defined by
\be \label{e:pots}
\frac{1}{r}\left[\frac{\partial\, \Lambda}{\partial \theta}-
\frac{1}{\sin \theta}\,\frac{\partial \Upsilon}{\partial \phi}\right]=
\hat{S}_{\theta},
\ \ \ \frac{1}{r}\left(\frac{1}{\sin \theta}\,\frac{\partial \Lambda}
{\partial \phi}+\frac{\partial \Upsilon}{\partial \theta}\right)=\hat{S}_{\phi}
\ee
\be\label{e:potg}
\frac{1}{r}\left(\frac{\partial\,U}{\partial \theta}-
\frac{1}{\sin \theta}\,\frac{\partial\,W}{\partial \phi}\right)=
V_{\theta},
\ \ \ \frac{1}{r}\left(\frac{1}{\sin \theta}\,\frac{\partial U}
{\partial \phi}+\frac{\partial W}{\partial \theta}\right)=V_{\phi}
\ee
allows to decouple Eqs.~(\ref{e:telgc-2}) and (\ref{e:telgc-3}).

Assume that the functions $\Lambda$ and $\Upsilon$ are known by means
of the resolution of the system (\ref{e:pots},\ref{e:potg}). After some
algebra, a solution of the telegraph equation (\ref{e:telgr}) can be
obtained by solving the associated equations
\begin{eqnarray}\label{e:telas}
D_{t} U+\tilde{\Delta} U &=& \Lambda \nonumber \\
D_{t} W+\tilde{\Delta} W &=& \Upsilon \ , 
\end{eqnarray}
where
\be\label{e:lapti}
\tilde{\Delta} \equiv \frac{\partial^{2}}{\partial r^{2}}+
\frac{1}{r^{2}}
\left(\frac{\partial{^2}}{\partial\,\theta^{2}}
+\frac{\cos \theta}{\sin \theta}\frac{\partial}{\partial \theta}+
\frac{1}{\sin \theta^{2}}\,\frac{\partial^{2}}{\partial \phi^{2}}\right)
\ee

It remains now to find the sources $\Lambda$ and $\Upsilon$ of the
system (\ref{e:telas}). Taking the angular divergence of
$\hat{S}_{\theta}$ and $\hat{S}_{\phi}$ allows to transform the system
(\ref{e:angpot}) as
\be\label{e:lapan}
\Delta_{\theta,\phi}\,\Lambda=\frac{\partial\hat{S}_{\theta}}{\partial \theta}
+\frac{\cos \theta}{\sin \theta}\,\hat{S}_{\theta}+
\frac{1}{\sin \theta^{2}}\frac{\partial\hat{S}_{\phi}}{\partial \phi} \ , 
\ee
where $\Delta_{\theta,\phi}$ is the angular part of the ordinary
Laplacian. This last equation can be easily solved by means of a
Fourier-Legendre transformation. Once $\Lambda$ is known, $\Upsilon$
can be easily computed.

It is to be noticed that the use of Cartesian components
$V_{x},\ V_{y},\ V_{z}$ would greatly simplify the resolution of the
telegraph equation. The drawback of this short cut is that only some
simple classes of BC can be easily implemented (e.g.
$V_{r}\,=\,V_{\theta}\,=\,V_{\phi}=0$). For more general BC, spherical
components of $\vec{V}$ and $\vec{S}$ must be used. Finally, let us
mention that the above method can be used to solve the vectorial
Poisson equation too, since it is a special case of Eq.~(\ref{e:telgr}),
obtained by setting $\alpha$, $\beta$ and $\lambda$ to zero. 
This integration method was used to solve a
vectorial Poisson equation in studying the bifurcation points of
general relativistic rotating neutron stars \cite{BonFG98}
(cf. Sect.~\ref{s:sym_break} below). 

\section{Astrophysical applications}
\label{s:astro}

\subsection{Einstein equations}\label{s:eis}

In this section we give for completeness a short outline of the
Einstein equations \cite{SmaY78}, \cite{Yor72}, \cite{Yor73}. This
section is not strictly necessary to understand the following
ones, and the reader that does not have some background of
differential geometry can skip it.

Einstein equations describe
the evolution of the metric $g_{\alpha\beta}$ of a 4-D Riemannian
manifold (Greek indices $\alpha,\beta$ run from $0$ to $3$, Latin
indices $i,j$ run from $1$ to $3$, $x^{0}=ct$). It is convenient to
write the distance element $ds^{2}$ on the following form:
\be\label{e:metr}
ds^{2} = -\alpha^2 (dx^0)^2 + \gamma_{ij} (dx^i - \beta^i dt)
	(dx^j - \beta^j dt) \ .  
\ee
The 10 Einstein equations (\ref{e:Einstein}) are not independent
because of the Bianchi identities (\ref{e:Bianchi}).  Therefore there
is the possibility to impose $4$ gauge conditions. If the {\em maximum
slicing} and {\em minimal distortion} gauge is chosen (cf.
\cite{SmaY78}), the Einstein equations (\ref{e:Einstein}) can be
written in the following symbolic way (where we have introduce the
Cartesian components of $\gamma_{ij}$)
\begin{eqnarray}
&&\Delta \alpha = \rho +Q_{0} \label{e:eiss1} \\ 
&&\Delta \beta_{i}+\frac{1}{3}\nabla_{i}\nabla_{j}\beta^{j}=S_{i}+Q_{i}
		\label{e:eiss2} \\
&&\Delta \log \gamma = P+\bar{Q} \label{e:eiss3}\\
&&\frac{1}{c^{2}}\frac{\partial{^2}\gamma_{ik}}{\partial t^{2}}\,
-\Delta \gamma_{ik}=P_{ik}+Q_{ik} \ , \label{e:eiss4}
\end{eqnarray}
where $\gamma$ is the determinant of the space metric $\gamma_{ik}$,
$\rho,\ S_{i},\ P$ and  $P_{ik}$  are related to the matter energy
momentum tensor $T_{\alpha \beta}$, $\Delta$ is the ordinary flat-space
Laplacian and $Q_{0},\ Q_{i},\ \bar{Q}$ and  $Q_{ik}$ represent a
formal source containing all  the non-linear terms  of the Einstein
equations. It is easy to see that the first five equations are
elliptical, the other ones are hyperbolic. The above equations can
simplify enormously if some symmetries are present. For example, in
spherical symmetry, a non-static solution of the system
(\ref{e:eiss1})-(\ref{e:eiss4}) can be found by solving a system of two
PDEs for the metric and two PDEs for the matter. An axial symmetric
circular steady state solution can be found by solving a systems of $4$
elliptical equations \cite{BonGSM93}. The problems becomes more
complicated in the noncircular case. In this case, in spite of the
axial symmetry, all the 10 equations of the system must be solved
\cite{GouB93}. In the next sections we shall describe some
astrophysical applications obtained by our group by solving the
Einstein equations (\ref{e:eiss1})-(\ref{e:eiss4}).

\subsection{1-D gravitational collapse of a neutron star towards a black hole}
\label{s:eff1d}

\begin{figure}
\centerline{\epsfig{file=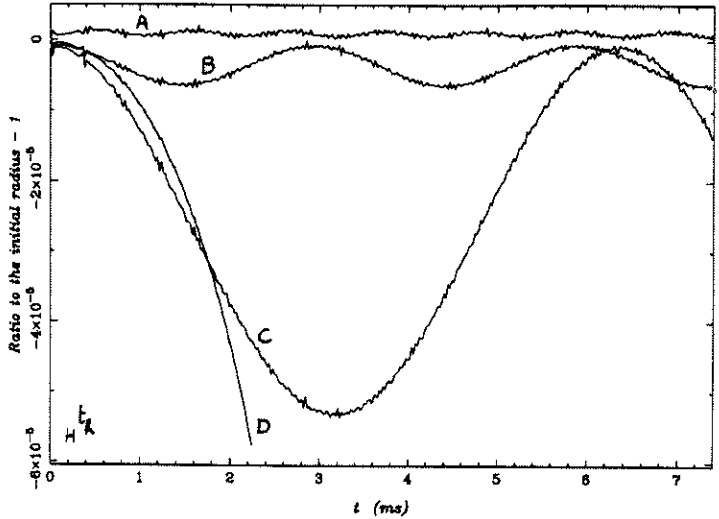,height=7cm}}
\caption{\label{f:oscillations} Time evolution of the stellar radius for
four initial configurations close to the neutron star maximum mass
\protect\cite{Gou92b}. The relative differences
with the maximum mass are respectively $-3.3\times 10^{-3}$,
$-5.0\times 10^{-5}$, $-3.6\times 10^{-6}$ and $-3.8\times 10^{-6}$
for configurations labelled A, B, C and D. The models A, B and C are
on the stable branch of the mass-central density curve, 
whereas model D is on the unstable branch. Note that the high frequency 
small oscillations are real and corresponds to stable acoustic modes. 
They keep the same amplitude and same frequency for the four configurations. 
On the contrary, the real part of the frequency of the unstable Jeans mode
tends to zero along the sequence A, B, C and its amplitude increases. 
For configuration D the frequency becomes imaginary, leading to a collapse
(not shown on the figure for the benefit of presentation).}
\end{figure}

Accreting neutron stars (e.g. in X-ray binary systems) can reach the
maximum mass allowed by general relativity \cite{OppV39}, which marks
the stability limit against radial perturbations \cite{HarTWW65}. Any
subsequent accretion will trigger the collapse towards a black hole.
This phenomenon may be an important source of neutrinos \cite{GouH93}.
We have studied the collapse by means of a general relativistic and
spherically symmetric hydrodynamical code \cite{Gou91}, \cite{Gou92b}.
The initial configurations were solutions of the hydrostatic equations
(Tolmann-Oppenheimer-Volkoff equations). The high accuracy of spectral
methods allows to take initial stable configurations, the mass of which
differs from the critical mass by a few $10^{-6}$ solar mass. The
numerical round-off error cause the solution not to remain exactly
static. It performs instead gentle oscillations of relative amplitude
$10^{-10}$ except for the fundamental mode, which grows exponentially
for initial unstable configurations (cf. Fig~\ref{f:oscillations}). The
collapse is then computed up to the formation of an apparent horizon
which leads to a coordinate singularity \cite{Gou91}.

\subsection{1-D gravitational collapse within tensor-scalar gravitational theory}
\label{s:tens_scal}

We have recently studied the gravitational wave emission from 1-D
gravitational collapse by solving the complete tensor-scalar and
hydrodynamic equations for a self-gravitating perfect fluid
\cite{Nov98b,Nov98c}. The initial conditions describe unstable-equilibrium
neutron star configuration. In the case of shock formation, we have
coupled our Poisson code to a Godunov-type solver for the hydro given
by Valencia's group \cite{Mar97}, \cite{Iba98}, \cite{Nov98b}. We
found that these kind of sources are not likely to be observed by
future laser interferometric gravitational wave detectors (such as
VIRGO or LIGO) if they are located at more than a few $100$ kpc.
However, spontaneous scalarization could be constrained if such a
gravitational collapse is detected by its quadrupolar gravitational
signal since this latter is quite lower than the monopolar one.

\subsection{2-D Kerr black hole} \label{s:Kerr}

\begin{figure}
\centerline{\epsfig{file=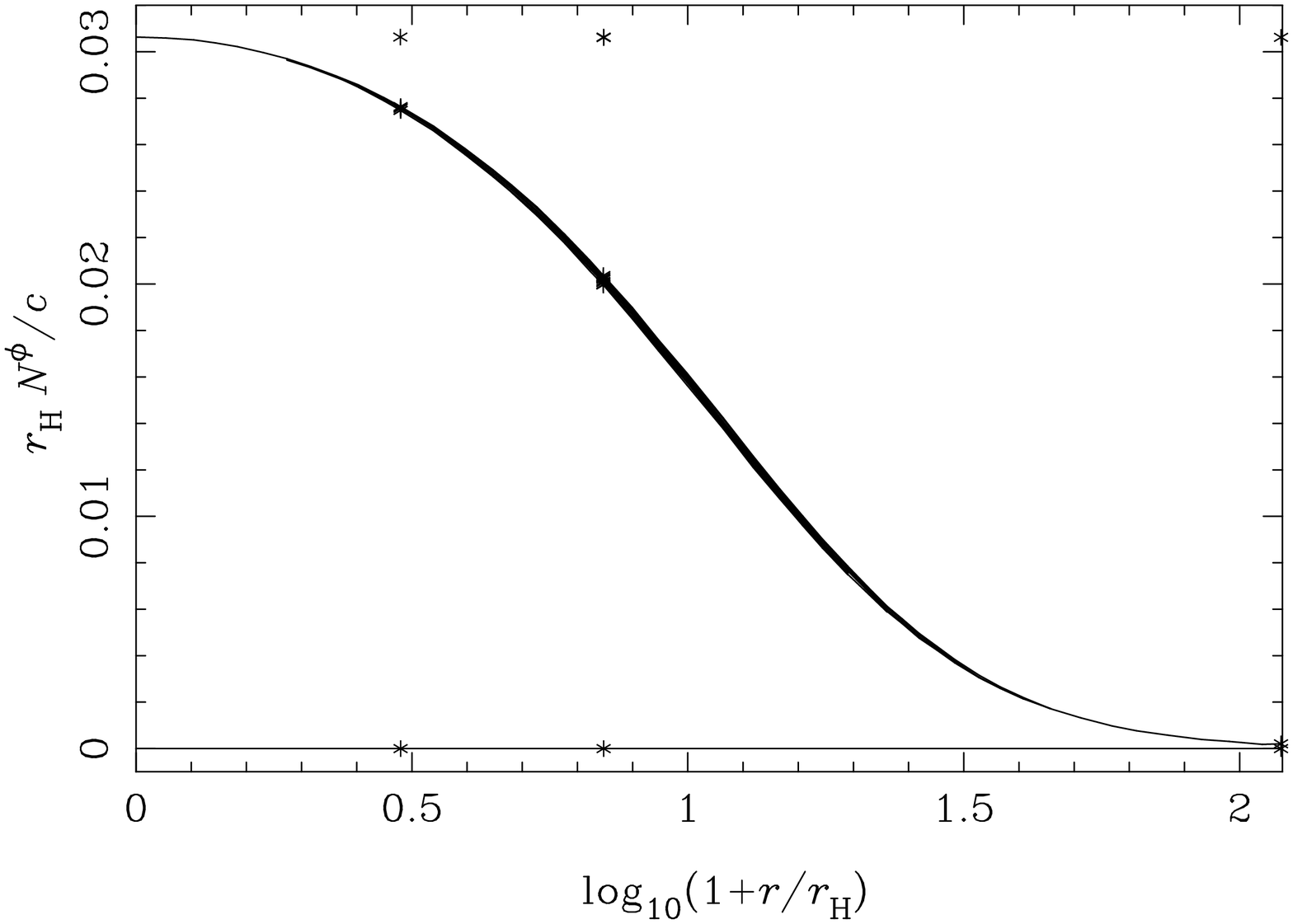,height=6cm}}
\centerline{\epsfig{file=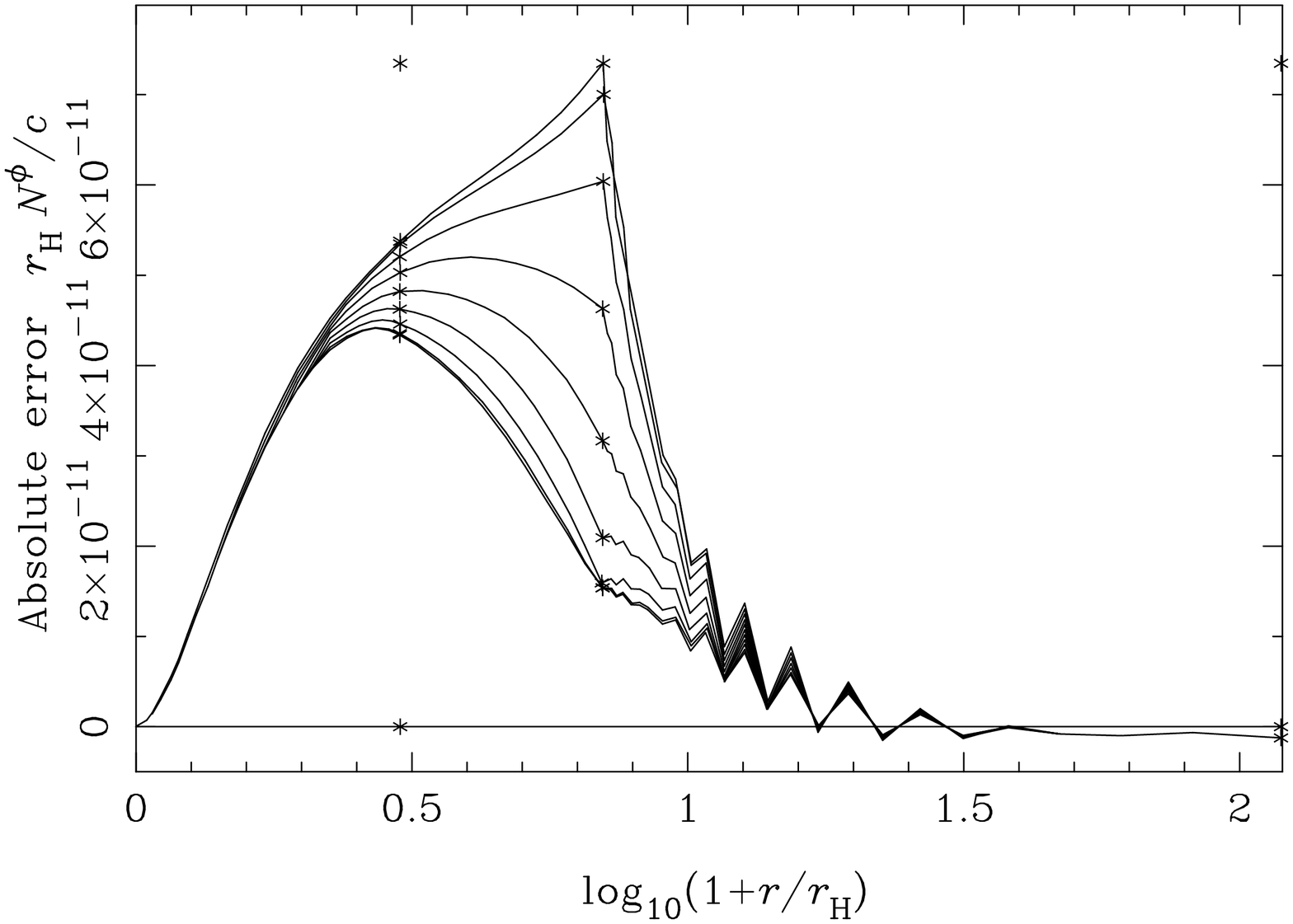,height=6cm}}
\caption{\label{f:shift-kerr} Shift vector component $\beta^\phi$
(top) and error with respect to the analytical solution (bottom)
for a Kerr black hole with angular momentum parameter $a/M=0.99$
\protect\cite{BonFG98}.  
The different curves corresponds to different values of the $\theta$ angle.
The asterisks indicate sub-domain boundaries.}
\end{figure}

We have tested our method for solving the axisymmetric stationary
Einstein equations against a non trivial 2-D analytical solution,
namely the Kerr solution (see e.g. \cite{Wal84}), which describes
rotating black holes.  For treating this problem, four equations must
be solved: Eq.~(\ref{e:eiss1}) for $\alpha$, only one of the three
Eqs.~(\ref{e:eiss2}) (that for $\beta^\phi$) and two equations
equivalent to the system (\ref{e:eiss3})-(\ref{e:eiss4}). These
equations are solved outside the event horizon (given by the
(isotropic) radial coordinate $r=1$) up to infinity, by means of a
compactification as explained in Sect.~\ref{s:posc}
\cite{BonFGM96,BonFG98}.  The horizon is a singular point for the
coefficient of the operator $\partial\over\partial r$ in
Eq.~(\ref{e:eiss2}). This problem is a typical Von Neumann-Dirichlet
problem. Figure~\ref{f:shift-kerr} shows the error on the shift
$\beta^\phi$ for a Kerr black hole with an angular momentum parameter
$a/M = 0.99$. Note that for $a/M=1$ the horizon shrinks to a point, so
that the singularity becomes essential.

\subsection{2-D magnetized rotating neutron stars} \label{s:rotNS}

\begin{figure}
\centerline{\epsfig{file=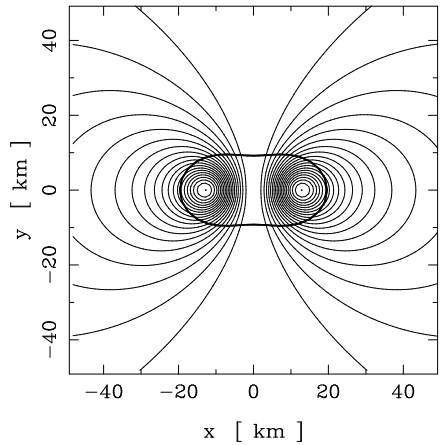,height=7cm}}
\caption{\label{f:cacahuete} Magnetic field lines and surface of the star 
(thick line) for a neutron star deformed by a huge magnetic pressure
\protect\cite{BocBGN95}.}
\end{figure}

\begin{figure}
\centerline{\epsfig{file=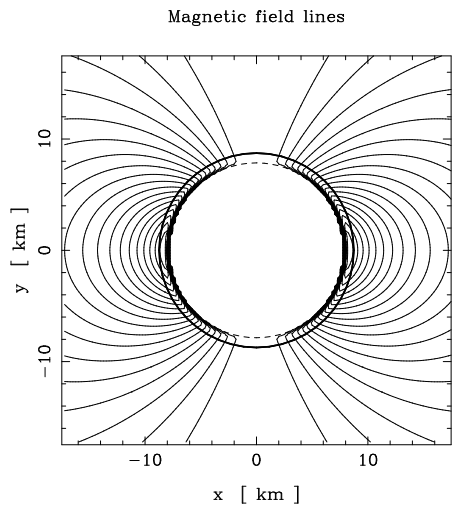,height=7cm}}
\caption{\label{f:magsupra} Magnetic field lines and surface of the star (thick
line) for a realistic magnetized neutron star, with superconducting interior
\protect\cite{BonG96a}.}
\end{figure}

We have developed a numerical code for solving the coupled
Einstein-Maxwell equations describing steady states of magnetized
rotating neutron stars. From the mathematical point of the view, this
problem amounts to solving four non-linear elliptical gravitational
field equations, as in Sect.~\ref{s:Kerr}, one elliptic equation for
the magnetic vector potential (the Grad-Shafranov equation) and one
elliptic equation for the electric scalar potential. These latter two
equations are coupled to the gravitational field ones via the metric
tensor. The steady-state configuration of the fluid is obtained by
using a first integral of motion \cite{BocBGN95}. This code has been
applied to compute the magnetic field induced distortion of rotating
neutron stars (Figs.~\ref{f:cacahuete} and \ref{f:magsupra}) in order
to estimate the resulting gravitational radiation \cite{BonG96a}.
Note that a direct comparison study has been conducted between the spectral
code and 
two codes based on different methods (finite differences) in the case
of non-magnetized rapidly rotating neutron stars \cite{NozSGE98}. 

\subsection{2-D hot new born neutron star}

We have computed models of differentially rotating proto-neutron stars
using realistic equations of state of dense hot matter in the framework
of general relativity \cite{GouHZ97} \cite{GouHZ98}. We found a minimum
period of uniform rotation of cooled neutron stars  
which formed directly (i.e. without a subsequent
significant accretion of mass) from proto-neutron stars with shocked
envelope of about 1.7~ms.  This strengthens the hypothesis that
millisecond pulsars are accretion accelerated neutron stars.

\subsection{3-D stellar core collapse}

A fully 3-D hydrodynamical Newtonian code for self-gravitating bodies
has been developed within spectral methods and employed to compute the
gravitational wave emission from the infall phase of a type II
supernova \cite{BonM93}. The initial conditions are self-consistent
configurations of rotating stellar cores embedded in the tidal field of
a companion, hence are fully triaxial. They are computed by means of an
iterative procedure. 

\subsection{3-D tidal disruption of a star by a massive black hole}

\begin{figure}
\centerline{\epsfig{file=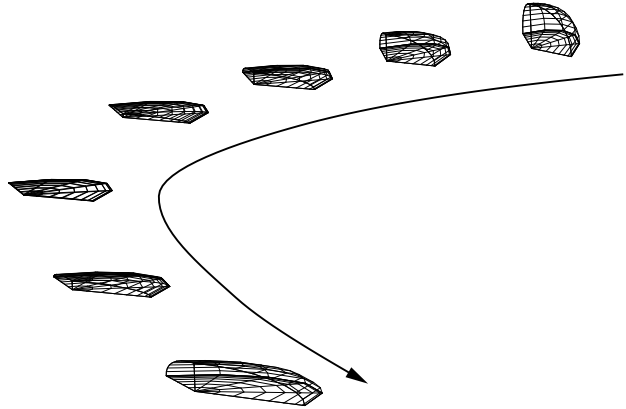,height=7cm}}
\caption{\label{f:crepe} Shape of the grid (adapted to the star) at different 
times during a 
close encounter with a massive black hole (the grid is initially the fourth of
a sphere)
\protect\cite{MarLB96}. 
This computation has been performed with only 17 degrees of freedom
in $r$, 9 in $\theta$ and 8 in $\phi$.}
\end{figure}

\begin{figure}
\centerline{\epsfig{file=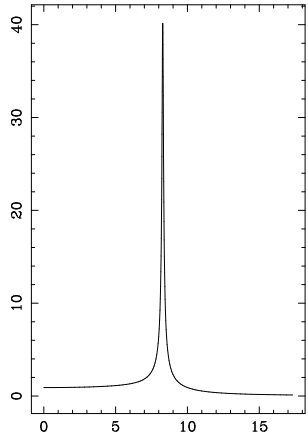,height=7cm}}
\caption{\label{f:compression} Time evolution of the central density of star
during its trip around a massive black hole \protect\cite{MarLB96}. }
\end{figure}

We have studied the close encounter of a star and massive black hole,
possibly at the centre of a galaxy \cite{MarLB96}. Tidal forces
compress dramatically the star, as shown in Figs.~\ref{f:crepe} and
\ref{f:compression}.  This phenomenon has been proposed as a possible
mechanism for gamma-ray burst generation \cite{Car92}.  The 3-D Euler
equations are solved in the external field of the black hole.  As
already said in Sect.~\ref{s:intro}, this application demonstrates the
capability of SM in computational resource saving.

\subsection{Spontaneous symmetry breaking of a rapidly rotating neutron
star} \label{s:sym_break}

We have computed the location of the bifurcation point between
(axisymme\-tric) MacLaurin-like and (triaxial) Jacobi-like configurations
along sequences of rapidly rotating stars in the framework of general
relativity \cite{BonFG96a,BonFG98}. To tackle this problem, we solved
Eq.~(\ref{e:eiss1}), the three Eqs.~(\ref{e:eiss2}) (by means of the
technique described in Sect.~\ref{s:telgr}), and two equations like
(\ref{e:eiss2}) and (\ref{e:eiss4}). The gravitational radiation is
neglected. The matter distribution is obtained thanks to a first
integral of motion and depends on the equation of state of nuclear matter.
Such mechanism can be a powerful source of continuous 
gravitational waves from accreting neutron stars in binary systems
\cite{BonG97a,BonG97b}.

\begin{figure}
\centerline{\epsfig{file=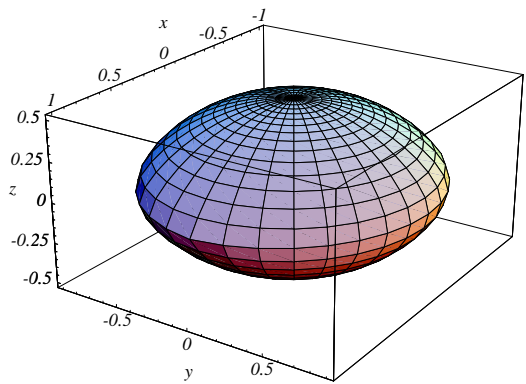,height=7cm}}
\caption{\label{f:triax} Rapidly rotating neutron star, which has become 
triaxial due to the viscosity driven instability \protect\cite{BonFG96a,BonFG98}.
The ellipticity in the 
equatorial plane is about 0.1. 
Notice the cusp at the stellar equator in direction
of the semi-major axis ($x$--axis), where the rotation is Keplerian,
and which is absent along the semi-minor axis ($y$--axis).}
\end{figure}

\subsection{Close binary systems of compact objects}

\begin{figure}
\centerline{\epsfig{file=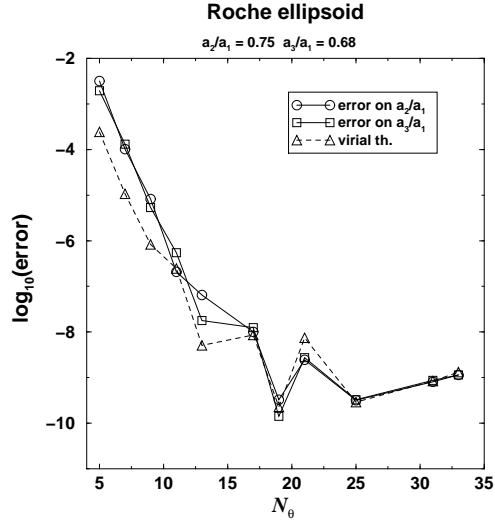,height=7cm}}
\caption{\label{f:Roche} 
Logarithm of the relative global error of the numerical solution
with respect to the number of degrees of freedom in
$\theta$ for a Roche ellipsoid for an equal mass binary
system and $\Omega^2/(\pi G\rho) = 0.1147$ 
(the numbers of degrees of freedom in the
other directions are $N_r = 2N_\theta-1$ and $N_\varphi = N_\theta -1$) 
\protect\cite{BonGM98a}.
Also shown is the error in the
verification of the virial theorem.}
\end{figure}

\begin{figure}
\centerline{\epsfig{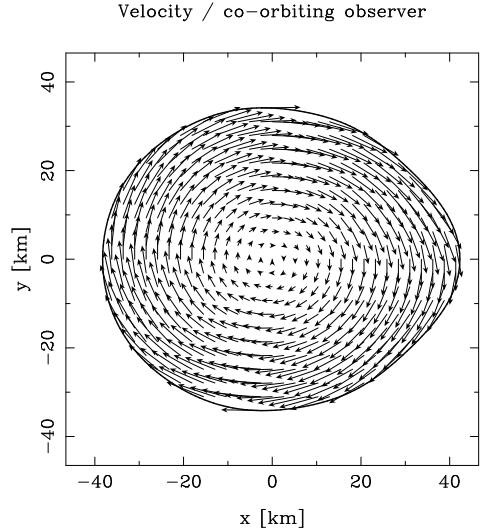}}
\caption{\label{f:vit_bin} 
Velocity field with respect to the orbiting frame, inside a star which
is member of a close binary system. The plane of the figure is the orbital
plane. The companion is located at the right.}
\end{figure}

We have developed a numerical code for computing quasi-equilibrium
configurations of relativistic binary systems \cite{BonGM98a}.  This
code is based on a {\sl multi-domain spectral method}, i.e. the whole
space is divided into various domains and a spectral method is used in
each domain, as described in Sect.~\ref{s:poisson-scal}. The external
domain extends to infinity thanks to some compactification.  The
boundaries of each domain are chosen in order to coincide with physical
limit surfaces, such as the surface of a star (cf.
Sect.~\ref{s:non_spher_domain}). In this manner, the physical
discontinuities in some fields or their derivatives are located at the
boundary of the domains and the applied spectral methods are free of
any Gibbs-phenomenon, resulting in a very high precision.  We use a
spherical-type coordinate system $(r',\theta',\phi')$ which maps the
physical domain to the unit sphere.  This technique is able to treat
any physical boundary which is starlike; this includes any realistic
shape taken by the surface of a rotating neutron star in a tidal
field.  Among the tests passed by the code, the comparison with
analytic solutions (MacLaurin and Roche ellipsoids) shows that the
achieved precision is of the order $10^{-10}$ (cf.
Figure~\ref{f:Roche}). The final stages of NS binaries are expected to
be irrotational with respect to an inertial frame (see \cite{BonGM97b}
and references therein). This means that when seen in the co-orbiting
frame, each star is counter-rotating with respect to the orbital
motion.  The velocity field with respect to the orbiting frame must
then to be computed.  A typical numerical solution is shown in
Fig.~\ref{f:vit_bin}. The first numerical results about irrotational
binary configurations in general relativity have been presented in
Ref.~\cite{BonGM98c}.

%
%
%
%

\section{Conclusion}

We hope to have convinced the reader that
spectral methods are a highly valuable tool in the field of
relativistic astrophysics. They can lead to results typically several orders of
magnitude more accurate than those provided by finite difference methods.
This high accuracy is very useful for studying stability problems
(Sect.~\ref{s:eff1d} and \ref{s:sym_break}). Of course, spectral
methods have their drawbacks. Let us summarize the advantages and drawbacks
from our experience:
\begin{itemize}
\item {\bf Advantages:} \begin{itemize}
	\item very high accuracy for treating smooth fields and their derivatives
		(evanescent error);
	\item economy of number of degrees of freedom (grid points);
	\item rigorous treatment of boundary conditions;
	\item multi-domain (multi-grid) technique easily implemented without
		any loss of accuracy;
	\item rigorous treatment of regularity conditions associated with
		singular coordinates such as spherical coordinates;
	\item efficient algorithms for von Neumann-Dirichlet problems;
	\item modularity;
	\item well developed mathematical theory;
	\item lack of robustness (this is an advantage because no solution 
	can be found for ill-posed problems);
	\end{itemize} 
\item {\bf Drawbacks:} \begin{itemize}
	\item spurious oscillations when treating discontinuous fields
	(Gibbs phenomenon);
	\item severe Courant conditions and therefore need of implicitation
		(Sect.~\ref{s:Tempor});
	\end{itemize}
	Note however that in the 2-D Fourier case, spectral hyperviscosity
	has proved to be efficient in shock handling \cite{PasP88,PasPPS90}.
\end{itemize}

Clearly there is no ideal numerical method for treating all the problems
and a combination of different
methods for solving a given problem may turn out to be fruitful, as 
preliminary results presented in Sect.~\ref{s:tens_scal} indicate.
It would also be desirable to compare results obtained by spectral methods
with that obtained by other methods, as was done for the problem of
rotating neutron stars (cf. Sect.~\ref{s:rotNS}) and 
extensively for hydrodynamical shocks \cite{WooC84}.

In our opinion, one of the main advantage of spectral methods is that
they allow a rigorous treatment of the regularity conditions associated
with the singularities of spherical coordinates. This is important because
spherical-like coordinates are obviously much more adapted than Cartesian
coordinates for describing objects like stars or black holes.
In particular, the number of points required to treat a star with Cartesian
coordinates with a reasonable accuracy is considerably higher than
that required by spherical coordinates. Moreover, the surface
of the star (or the black hole horizon), where boundary conditions may
have to be set, is described more simply with spherical coordinates, 
especially when surface-fitted spherical coordinates are introduced as in 
Sect.~\ref{s:non_spher_domain}.  
Also the boundary conditions for Poisson-type equations are better expressed 
in spherical coordinates than in Cartesian ones. 
 
\bigskip
We thank Prof.~Jos\'e-Maria Ib\'a\~nez and an anonymous referee for their 
carefull reading of the manuscript.

\end{document}